\documentclass[12pt,a4paper]{article}

\title{Pleba\'nski--Demia\'nski solution of general relativity and its expressions quadratic and cubic in curvature: analogies to electromagnetism}

\author{
	Jens Boos\footnote{E-mail: \href{mailto:boos@thp.uni-koeln.de}{boos@thp.uni-koeln.de}} \\
	{\small Institute for Theoretical Physics}\\[-5pt]
	{\small University of Cologne, 50937 K\"oln, Germany} \\
}

\date{}

\usepackage[english]{babel}
\usepackage{amsmath}
\usepackage{amsbsy}
\usepackage{amstext}
\usepackage{amscd}
\usepackage{amsxtra}
\usepackage{amsopn}
\usepackage[onehalfspacing]{setspace}
\usepackage{tensor}
\usepackage{mathrsfs}
\usepackage{relsize}
\usepackage{amsfonts}
\usepackage{MnSymbol}
\usepackage[makeroom]{cancel}
\usepackage{selinput}
\usepackage{bbm}

\usepackage{geometry}
\usepackage{parskip}

\usepackage[colorlinks=true,citecolor=green,linkcolor=red,filecolor=cyan,urlcolor=magenta,backref=page]{hyperref}
\renewcommand*{\backref}[1]{}
\renewcommand*{\backrefalt}[4]{%
    \ifcase #1 (Not cited.)%
    \or        (Cited on page~#2.)%
    \else      (Cited on pages~#2.)%
    \fi}

\usepackage{color}
\definecolor{mygray}{rgb}{0.5,0.5,0.5}

\usepackage{listings}
\newcommand{\tvect}[2]{%
  \ensuremath{\Bigl(\negthinspace\begin{smallmatrix}#1\\#2\end{smallmatrix}\negthinspace\Bigr)}}

\definecolor{fadedred}{rgb}{0.6,0.0,0.0}
\definecolor{fadedblue}{rgb}{0.0,0.0,0.6}

\begin{document}

\maketitle

\begin{abstract}
	Analogies between gravitation and electromagnetism have been known since the 1950s. Here, we examine a fairly general type D solution---the exact seven parameter solution of Pleba\'nski--Demia\'nski (PD)---to demonstrate these analogies for a physically meaningful spacetime:

	The two quadratic curvature invariants $\bf{B}^2-\bf{E}^2$ and $\bf{E}\cdot\bf{B}$ are evaluated analytically. In the asymptotically flat case, the leading terms of $\bf{E}$ and $\bf{B}$ can be interpreted as gravitoelectric mass and gravitoelectric current of the PD solution, respectively, if there are no gravitomagnetic monopoles present.

	Furthermore, the square of the Bel--Robinson tensor reads $(\mathbf{B}^2+\mathbf{E}^2)^2$ for the PD solution, reminiscent of the square of the energy density in electrodynamics. By analogy to the energy-momentum 3-form of the electromagnetic field, we provide an alternative way to derive the recently introduced Bel--Robinson 3-form, from which the Bel--Robinson tensor can be calculated.

	We also determine the Kummer tensor, a tensor cubic in curvature, for a general type D solution for the first time, and calculate the pieces of its irreducible decomposition.

	The calculations are carried out in two coordinate systems: in the original polynomial PD coordinates, and in a modified Boyer--Lindquist-like version introduced by Griffiths and Podolsk\'y (GP) allowing for a more straightforward physical interpretation of the free parameters. {\small \textit{file: 05\_elm\_inv\_v9.tex, Jul 2, 2015} }
\end{abstract}

\vspace{7em}

\pagebreak

\section*{Introduction}

It has long been known that general relativity and electrodynamics exhibit certain formal similarities, see Matte \cite{Matte:1953} and Bel \cite{Bel:1962}, or---for a more recent overview---Costa and Nat\'ario \cite{Costa:2012cw}. On the other hand, Petrov type D spacetimes can be thought of as spacetimes with Coulomb-like sources, as already pointed out by Szekeres \cite{Szekeres:1965}; see also Stephani \textit{et al}.\ \cite{Stephani:2003}, p.\ 39. In this sense, they are the closest scenario to electro/magneto-statics that general relativity has to offer.

The Pleba\'nski--Demia\'nski solution \cite{Plebanski:1976gy} is of type D, has seven free parameters, and can be used to model various axisymmetric electrovacuum spacetimes, in particular the Kerr solution \cite{Kerr:1963ud}. Overviews of original papers for various subclasses of the PD solution can be found in Pleba\'nski and Demia\'nski \cite{Plebanski:1976gy}, Table 1, or Stephani \textit{et al}.\ \cite{Stephani:2003}, Table 21.1.

For our work, we found particularly useful the reviews on Kerr--Taub--NUT--(A)de\,Sitter by Mars and Senovilla \cite{Mars:2013qja}; the (rotating) $C$-metric by Hong and Teo \cite{Hong:2003gx, Hong:2004dm}; Kerr-de\,Sitter by Chambers \cite{Chambers:1994ap}; and of course the Kerr metric itself, see for example Carter \cite{Carter:1973}, de\,Felice and Bradley \cite{deFelice:1988uk}, and Cherubini \textit{et al}.\ \cite{Cherubini:2003nj}.

A complete physical interpretation of the PD coordinates and its free parameters has been lacking for a long time, but it is possible to transform the PD coordinates to Boyer--Lindquist-like coordinates, see Griffiths and Podolsk\'y \cite{Griffiths:2005qp}.

The purpose of this paper is twofold: First of all, it is of general interest to calculate various geometric quantities of the PD solution in physically well-motivated coordinates. And secondly, the setting of the PD solution may endow the more general, formal similarities between gravitation and electrodynamics with an explicit physical understanding.

We proceed as follows: In the first section, we fix our notation and demonstrate that the PD solution satisfies the Einstein--Maxwell equations. In the second section, we briefly check the coordinate transformations introduced by Griffiths and Podolsk\'y, and confirm that the PD solution, expressed in their Boyer--Lindquist-like coordinates, still fulfills the Einstein--Maxwell equations. In the third section, we decompose the curvature irreducibly and calculate its pieces for the PD solution in both coordinates. For the asymptotically flat case, the asymptotic structure of the Weyl 2-form is analyzed and compared to classical electrodynamics. In the fourth section, we move on to the quadratic curvature invariants and recover the well-known analogies between general relativity and electrodynamics. Drawing on these analogies, we provide a quick way to derive the recently introduced Bel--Robinson 3-form in section 5 and confirm its correspondence to the Bel--Robinson tensor. Finally, we turn to cubic expressions in curvature and calculate the pieces of the irreducible decomposition of the Kummer tensor, an important tensor related to the principal null directions of curvature, for a general type D spacetime. In the concluding section, we briefly summarize our findings.

\section{Pleba\'nski--Demia\'nski solution and its parameters}
The Pleba\'nski--Demia\'nski solution of 1976 is an exact solution of the Einstein--Maxwell equations with cosmological constant \cite{Plebanski:1976gy}. Using the coordinates $\{\tau, p, q, \sigma\}$ it can be written in terms of the pseudo-orthonormal coframe
\begin{align}
\begin{split}
\tensor{\vartheta}{^{\hat{0}}} & := \frac{1}{1 - pq} \sqrt{\frac{\mathscr{Q}(q)}{p^2 + q^2}}\left( d\tau - p^2d\sigma \right) , \\
\tensor{\vartheta}{^{\hat{1}}} & := \frac{1}{1 - pq} \sqrt{\frac{p^2 + q^2}{\mathscr{Q}(q)}} \, dq , \\
\tensor{\vartheta}{^{\hat{2}}} & := \ominus \, \frac{1}{1 - pq} \sqrt{\frac{p^2 + q^2}{\mathscr{P}(p)}} \, dp , \\
\tensor{\vartheta}{^{\hat{3}}} & := \ominus \, \frac{1}{1 - pq} \sqrt{\frac{\mathscr{P}(p)}{p^2 + q^2}} \left( d\tau + q^2d\sigma \right) . \label{eq:pb_coframe}
\end{split}
\intertext{The metric is $g = -\tensor{\vartheta}{^{\hat{0}}}\otimes\tensor{\vartheta}{^{\hat{0}}} + \tensor{\vartheta}{^{\hat{1}}}\otimes\tensor{\vartheta}{^{\hat{1}}} + \tensor{\vartheta}{^{\hat{2}}}\otimes\tensor{\vartheta}{^{\hat{2}}} + \tensor{\vartheta}{^{\hat{3}}}\otimes\tensor{\vartheta}{^{\hat{3}}}$, and hatted indices $\hat{0}, \hat{1}, \hat{2}, \hat{3}$ will be used to refer to the anholonomic components in this frame. $\ominus = \pm 1$ determines the spatial orientation of the angular parts of the coframe: for $\ominus = +1$, as chosen originally by PD, we would not retrieve the flat spatial tetrad in its usual orientation, as will be shown later. At this point, we merely include this symbol without fixing its value yet. The vector potential 1-form reads ($\hat{e} \sim$~electric charge, $\hat{g} \sim$ magnetic charge)}
\begin{split}
A & := \frac{1-pq}{\sqrt{p^2 + q^2}} \left( \frac{\hat{e}q}{\sqrt{\mathscr{Q}(q)}} \tensor{\vartheta}{^{\hat{0}}} + \frac{\hat{g}p}{\sqrt{\mathscr{P}(p)}} \tensor{\vartheta}{^{\hat{3}}} \right) . \label{eq:pb_potential}
\end{split}
\end{align}
The related field strength is $F := dA$ and the excitation (in vacuum) is given by $\mu_0H := \star F$. We use CGS units where $\{4\pi\epsilon_0 = 1, \mu_0 = 4\pi\}$; see Hehl and Obukhov \cite{Hehl:2003} for a detailed introduction to electrodynamics formulated in terms of exterior calculus. $\mathscr{P}$ and $\mathscr{Q}$ are quartic functions defined via
\begin{align}
\begin{split}
\mathscr{P}(p) & := \hat{k} + 2\hat{n}p - \hat{\epsilon} p^2 + 2\hat{m}p^3 + \left( \hat{k} + \hat{e}^2 + \hat{g}^2 - \frac{\Lambda}{3} \right)p^4 , \\
\mathscr{Q}(q) & := \hat{k} + \hat{e}^2 + \hat{g}^2 - 2\hat{m}q + \hat{\epsilon} q^2 - 2\hat{n}q^3 + \left( \hat{k} - \frac{\Lambda}{3} \right)q^4 . \label{eq:pb_quartics}
\end{split}
\end{align}
The identification $\hat{k} := \hat{\gamma} - \hat{g}^2 - \Lambda/6$ readily reproduces Eq.\ (3.31) given in \cite{Plebanski:1976gy}. Using computer algebra (see appendix \ref{appendix:computer_algebra}), it is quite straightforward to show that Eqs.\ (\ref{eq:pb_coframe})--(\ref{eq:pb_quartics}) indeed solve the Einstein--Maxwell equations.

In exterior calculus (see our conventions in appendix \ref{appendix:exterior_calculus}), the Einstein--Maxwell equations with cosmological constant $\Lambda$ take the form
\begin{align}
\frac{1}{2} \tensor{\eta}{_\mu_\alpha_\beta} \wedge \tensor{\text{Riem}}{^\alpha^\beta} + \Lambda \tensor{\eta}{_\mu} = 8\pi \tensor{\Sigma}{_\mu} . \label{eq:einstein_maxwell}
\end{align}
$\tensor{\Sigma}{_\mu}$ is the electromagnetic energy-momentum 3-form defined via
\begin{align}
\tensor{\Sigma}{_\mu} := \frac{1}{2} \left[ F \wedge \left( \tensor{e}{_\mu} \righthalfcup H \right) - H \wedge \left( \tensor{e}{_\mu} \righthalfcup F \right) \right] . \label{eq:electrodynamics_energy_momentum_3_form}
\end{align}
Evaluating the left-hand side of Eq.\ \eqref{eq:einstein_maxwell} explicitly yields
\begin{align}
\frac{1}{2} \tensor{\eta}{_\mu_\alpha_\beta} \wedge \tensor{\text{Riem}}{^\alpha^\beta} + \Lambda \tensor{\eta}{_\mu} = \frac{\hat{e}^2 + \hat{g}^2}{(p^2 + q^2)^2}(1-pq)^4 \, \tensor{\eta}{_\mu} .
\end{align}
The energy-momentum 3-form of the potential \eqref{eq:pb_potential} reads
\begin{align}
\tensor{\Sigma}{_\mu} = \frac{(1-pq)^4}{8\pi}\frac{\hat{e}^2 + \hat{g}^2}{(p^2 + q^2)^2} \, \tensor{\eta}{_\mu}.
\end{align}
Therefore, Eqs.\ \eqref{eq:pb_coframe}--\eqref{eq:pb_quartics} indeed solve the Einstein--Maxwell equations. Note that the trace $\Sigma := \tensor{\vartheta}{_\alpha} \wedge \tensor{\Sigma}{^\alpha}$ vanishes, as expected for an electromagnetic field. Furthermore, the constants $\hat{e}$ and $\hat{g}$ can tentatively be identified as the electric and magnetic charge, respectively.

\section{New coordinates of Griffiths and Podolsk\'y}
As seen above, the PD solution comes with a set of \emph{seven} free parameters: $\{\hat{m}, \hat{n}, \hat{e}, \hat{g}, \hat{\epsilon}, \hat{k}, \Lambda\}$. Provided that at least one of the parameters $\{\hat{m}, \hat{n}, \hat{e}, \hat{g}\}$ does not vanish, the solution is of Petrov type D. By means of the Einstein--Maxwell equation, $\Lambda$ can be identified straightforwardly as the cosmological constant. The parameters $\{\hat{e}, \hat{g}\}$ allow an interpretation as electric charge and magnetic charge, respectively, by their appearance in the energy-momentum 3-form. At any rate, the nature of the remaining four parameters remains somewhat obscure.

However, this solution contains a variety of limiting cases, such as Schwarzschild, Taub--NUT, Kerr(--Newman), de\,Sitter, the $C$-metric, and combinations thereof. It is the lack of direct physical meaning of the free parameters $\{\hat{m}, \hat{n}, \hat{\epsilon}, \hat{k}\}$ that makes it difficult to procure a simple limiting procedure to arrive at the aforementioned spacetimes. This has already been pointed out by Griffiths and Podolsk\'y \cite{Griffiths:2005qp}. At this point, we will briefly summarize their approach to extract a physically directly relevant coframe and vector potential.

Following Hong and Teo \cite{Hong:2003gx, Hong:2004dm}, a coordinate transformation can be employed to simplify the roots of the quartics $\mathscr{P}, \mathscr{Q}$ of Eq.\ \eqref{eq:pb_quartics}. This is a promising procedure, since the quartics control the Lorentzian signature of the metric.

The transformation introduces the new coordinates $\{t, r, \tilde{p}, \phi\}$, as well as the new parameters $\alpha$, $\omega$, $a$, and $\ell$. It is degenerate provided any of the new parameters (save $\ell$) vanish:
\begin{equation}
\begin{alignedat}{3}
\tau ~ & ~ \mapsto ~ \tau(t, \phi) && ~ := ~ \sqrt{\frac{\omega}{\alpha}}\left(t - \frac{(\ell + a)^2}{a}\phi \right) , \\
p ~ & ~ \mapsto ~ p(\tilde{p}) && ~ := ~  \sqrt{\frac{\alpha}{\omega}}\left( \ell + a\tilde{p} \right) , \\
q ~ & ~ \mapsto ~ q(r) && ~ := ~ \sqrt{\frac{\alpha}{\omega}} r , \\
\sigma ~ & ~ \mapsto ~ \sigma(\phi) && ~ := ~ - \, {\left(\frac{\omega}{\alpha}\right)}^\frac{3}{2} \, \frac{\phi}{a} . \label{eq:gp_coordinate_transformation}
\end{alignedat}
\end{equation}
Simultaneously, the free parameters of the original PD solution are scaled according to
\begin{equation}
\begin{alignedat}{3}
\left(\hat{m}, \hat{n} \right) ~ & ~ \mapsto ~ \left( \frac{\alpha}{\omega} \right)^{\frac{3}{2}} \left(m, n \right) , \\
\left(\hat{e}, \hat{g} \right) ~ & ~ \mapsto ~ \frac{\alpha}{\omega} \left(e, g \right) , \\
\hat{\epsilon} ~ & ~ \mapsto ~ \frac{\alpha}{\omega} \epsilon , \\
\hat{k} ~ & ~ \mapsto ~ \alpha^2 k , \\
\Lambda ~ & ~ \mapsto ~ \Lambda . \label{eq:gp_constant_redefinition}
\end{alignedat}
\end{equation}
Note that also this scaling is degenerate in the cases of vanishing $\omega$, $a$ or $\alpha$. With the additional parameters, there are 11 degrees of freedom. Three of these degrees of freedom can be used to adjust the roots of $\mathscr{P}$ to $\tilde{p} = \pm 1$, thereby introducing the relations
\begin{align}
\begin{split}
\epsilon & := \frac{\omega^2 k}{a^2 - \ell^2} + 4\frac{\alpha}{\omega}\ell m - (a^2 + 3\ell^2) \left[\frac{\alpha^2}{\omega^2}(\omega^2 k + e^2 + g^2) - \mathsmaller{\frac{\Lambda}{3}} \right] , \\
n & := \frac{\omega^2 k\ell}{a^2 - \ell^2} - \frac{\alpha}{\omega}(a^2 - \ell^2)m  + (a^2 - \ell^2) \ell \left[ \frac{\alpha^2}{\omega^2}(\omega^2k + e^2 + g^2) - \mathsmaller{\frac{\Lambda}{3}} \right] , \\
k & := \frac{(a^2 - \ell^2)(1 + 2\frac{\alpha}{\omega}\ell m - 3\frac{\alpha^2}{\omega^2}\ell^2(e^2 + g^2) + \ell^2\Lambda)}{\omega^2 + (a^2 - \ell^2)3\alpha^2\ell^2} . \label{eq:gp_constraints}
\end{split}
\end{align}
The parameters $\epsilon, n, k$ are now fixed by the new parameters $\alpha$, $\omega$, $a$, and $\ell$. According to \cite{Griffiths:2005qp}, the remaining degree of freedom can be used to set $\omega$ to a convenient value, if not both $a$ and $\ell$ vanish simultaneously. The quartics become
\begin{align}
\begin{split}
\mathscr{P} & = \frac{\alpha^2a^2}{\omega^2} \left(1 - \tilde{p}^2 \right) \left( 1 - \alpha_3\tilde{p} - \alpha_4\tilde{p}^2 \right) , \\
\mathscr{Q} & = \frac{\alpha^2}{\omega^2} \Big[ \omega^2 k + e^2 + g^2 -2mr + \epsilon r^2 - 2\mathsmaller{\frac{\alpha}{\omega}} nr^3 - \left( \alpha^2k - \mathsmaller{\frac{\Lambda}{3}} \right) r^4 \Big] .
\end{split}
\end{align}
The constants $\alpha_3$, $\alpha_4$ turn out to be
\begin{align}
\begin{split}
\alpha_3 & := 2\frac{\alpha}{\omega} a m - 4 a \ell \frac{\alpha^2}{\omega^2} \left( \omega^2 k + e^2 + g^2 \right) + 4a\ell \frac{\Lambda}{3} , \\
\alpha_4 & := -\frac{\alpha^2}{\omega^2}a^2 \left( \omega^2 k + e^2 + g^2 \right) + \frac{\Lambda}{3} a^2 . \label{eq:gp_constants}
\end{split}
\end{align}
Since on $-1 < \tilde{p} < 1$ due to $\mathscr{P} \ge 0$ we have a Lorentzian signature (assuming $\alpha_3$ and $\alpha_4$ are sufficiently small), a convenient parametrization is $\tilde{p} = \cos\theta$, with $\theta \in [0, \pi]$. We then arrive at the final expression for the PD coframe expressed in more familiar Boyer--Lindquist-like coordinates $\{t, r, \theta, \phi\}$:
\begin{align}
\begin{split}
\tensor{\vartheta}{^{\hat{0}}} & := \frac{\sqrt{\Delta}}{\Omega\rho} \left[ dt - \left( a\sin^2\theta +4\ell\sin^2 \mathsmaller{\frac{\theta}{2}} \right) \, d \phi \right] , \\
\tensor{\vartheta}{^{\hat{1}}} & := \frac{\rho}{\Omega\sqrt{\Delta}} \, dr , \\
\tensor{\vartheta}{^{\hat{2}}} & := \ominus \, \frac{-\rho}{\Omega\sqrt{\chi}} \, d\theta , \\ 
\tensor{\vartheta}{^{\hat{3}}} & := \ominus \, \frac{\sqrt{\chi}\sin\theta}{\Omega\rho}\left\{ a \, dt - \left[ r^2 + (a + \ell)^2 \right] \, d\phi  \right\} . \label{eq:gp_coframe}
\end{split}
\intertext{Note that it is now convenient to set $\ominus := -1$ such that the angular part of the coframe has its standard sign for flat spacetime. The vector potential reads}
A & := \frac{\Omega}{\rho}\left[ \frac{e r}{\sqrt{\Delta}} \, \tensor{\vartheta}{^{\hat{0}}} ~ + ~ \frac{g (\ell/a + \cos\theta)}{\sin\theta\sqrt{\chi}} \, \tensor{\vartheta}{^{\hat{3}}} \right] . \label{eq:gp_potential}
\end{align}
We introduced the following auxiliary functions:
\begin{align}
\begin{split}
\Delta &:= \omega^2k + e^2 + g^2 - 2mr + \epsilon r^2 - 2\frac{\alpha}{\omega} n r^3 - \left(\alpha^2 k - \frac{\Lambda}{3}\right)r^4 ~ = ~ \left(\frac{\omega}{\alpha}\right)^2 \mathscr{Q} , \\
\chi   &:= 1 - \alpha_3\cos\theta - \alpha_4\cos^2\theta ~ = ~ \frac{\omega^2}{\alpha^2 a^2 \sin^2\theta} \mathscr{P} , \\
\Omega &:= 1 - \frac{\alpha}{\omega} r (\ell + a\cos\theta) ~ = ~ 1 - pq , \\
\rho^2 &:= r^2 + (\ell + a\cos\theta)^2 ~ = ~ \frac{\omega}{\alpha}\left( p^2 + q^2 \right). \label{eq:gp_aux}
\end{split}
\end{align}
Simple computer algebra (see appendix \ref{appendix:computer_algebra}) can be used to verify that Eqs.~\eqref{eq:gp_coframe}--\eqref{eq:gp_aux} solve the Einstein--Maxwell equations for any value of $\omega$. The left-hand side of the Einstein--Maxwell equations turns out to be
\begin{align}
\frac{1}{2} \tensor{\eta}{_\mu_\alpha_\beta} \wedge \tensor{\text{Riem}}{^\alpha^\beta} + \Lambda\tensor{\eta}{_\mu} = \frac{\Omega^4(e^2 + g^2)}{\rho^4} \, \tensor{\eta}{_\mu} .
\end{align}
The energy-momentum 3-form derived from the potential \eqref{eq:gp_potential} reads
\begin{align}
\tensor{\Sigma}{_\mu} = \frac{1}{8\pi} \frac{\Omega^4(e^2 + g^2)}{\rho^4} \, \tensor{\eta}{_\mu} .
\end{align}
Thus, Eqs.\ \eqref{eq:gp_coframe}--\eqref{eq:gp_aux} indeed fulfill the Einstein--Maxwell equations for any $\omega$.

Griffiths and Podolsk\'y \cite{Griffiths:2005qp} interpret the new set of free parameters to be $\{m, \ell, a, \alpha, e, g, \Lambda\}$, combined with a scaling degree of freedom $\omega$. All of the parameters have a physical interpretation: mass, Taub--NUT parameter, angular momentum parameter, acceleration parameter, electric and magnetic charge, and cosmological constant, respectively. This can be seen, for example, by determining the coframes for various choices of the parameters and comparing it to the literature.

It is noteworthy, however, that the Einstein--Maxwell equations are fulfilled for any value of $\omega$. Therefore the interpretation of $\omega$ as a pure scaling degree of freedom becomes questionable. Similar results have been found for the original PD solution by Garc\'ia and Mac\'ias \cite{Garcia:1998}, see also Socorro \textit{et al}.\ \cite{Socorro:1998hr}. How their additional parameter $\mu$ is related to the $\omega$ parameter of Griffiths and Podolsk\'y will be subject of a further study. 

The seemingly divergent expressions $(a^2 - \ell^2)$ in Eqs.\ \eqref{eq:gp_constraints}$_1$, \eqref{eq:gp_constraints}$_2$ cancel the leading $(a^2 - \ell^2)$ of Eq.\ \eqref{eq:gp_constraints}$_3$. Therefore, all limits are well-behaved, even in the case $a = \ell = 0$. However, in this case, the parameter $\omega$ has to be adjusted appropriately. A compilation of spacetimes, somewhat similar as \cite{Griffiths:2005qp}, can be found in Table \ref{table:coframes}.

\begin{table} \centering
	\begin{tabular}{c|l}

		parameters & \hspace{8pt} coframe \\

		\hline \parbox{1.5cm}{ \centering $m$, $a$,\\$e$, $\Lambda$,\\$\omega = 1$ } &
		\parbox{13cm}{ \begin{flalign*} \hspace{10pt}
			& \text{\textbf{Kerr--Newman--de\,Sitter}} && \nonumber \\[10pt]
			\tensor{\vartheta}{^{\hat{0}}} &= \frac{\sqrt{r^2 - 2mr + a^2 + \frac{1}{3}\Lambda(r^4 + a^2r^2) + e^2}}{\sqrt{r^2 + a^2\cos^2\theta}} \left( dt - a\sin^2\theta d\phi \right) && \\
			\tensor{\vartheta}{^{\hat{1}}} &= \frac{\sqrt{r^2 + a^2\cos^2\theta}\,dr}{\sqrt{r^2 - 2mr + a^2 + \frac{1}{3}\Lambda(r^4 + a^2r^2) + e^2}} && \\
			\tensor{\vartheta}{^{\hat{2}}} &= \frac{\sqrt{r^2 + a^2\cos^2\theta} \, d\theta}{\sqrt{1 - \frac{1}{3}\Lambda a^2\cos^2\theta}} && \\
			\tensor{\vartheta}{^{\hat{3}}} &= \sqrt{\frac{1 - \frac{1}{3}\Lambda a^2\cos^2\theta}{r^2 + a^2\cos^2\theta}} \sin\theta \Big[(r^2 + a^2)d\phi - a dt \Big]  && \\
			A &= \frac{er \, \tensor{\vartheta}{^{\hat{0}}}}{\sqrt{r^2 - 2mr + a^2 + \frac{1}{3}\Lambda(r^4 + a^2r^2) + e^2}} &&
		\end{flalign*} } \\[15pt]
	
		\hline \parbox{1.5cm}{ \centering $m$, $\ell$,\\$e$, $\Lambda$,\\$\omega = 1$ } &
		\parbox{13cm}{ \begin{flalign*} \hspace{10pt}
			& \text{\textbf{charged Taub--NUT--de\,Sitter}} && \nonumber \\[10pt]
			\tensor{\vartheta}{^{\hat{0}}} &= \sqrt{\frac{r^2\left(1 + 2\Lambda\ell^2\right) + e^2 - 2mr - \ell^2\left(1 - \Lambda\ell^2\right) + \frac{1}{3}\Lambda r^4}{r^2 + \ell^2}} \Big[ dt - 2\ell \left(1 - \cos\theta \right) d\phi \Big] && \\
			\tensor{\vartheta}{^{\hat{1}}} &= \frac{\sqrt{r^2 + \ell^2} \, dr}{\sqrt{r^2\left(1 + 2\Lambda\ell^2\right) + e^2 - 2mr - \ell^2\left(1 - \Lambda\ell^2\right) + \frac{1}{3}\Lambda r^4}} && \\
			\tensor{\vartheta}{^{\hat{2}}} &= \sqrt{r^2 + \ell^2} \, d\theta && \\
			\tensor{\vartheta}{^{\hat{3}}} &= \sqrt{r^2 + \ell^2} \sin\theta \, d\phi && \\
			A &= \frac{er \, \tensor{\vartheta}{^{\hat{0}}}}{\sqrt{r^2\left(1 + 2\Lambda\ell^2\right) + e^2 - 2mr - \ell^2\left(1 - \Lambda\ell^2\right) + \frac{1}{3}\Lambda r^4}} &&
		\end{flalign*} } \\
	
		\hline \parbox{1.5cm}{ \centering $m$, $\alpha$,\\$\omega = a$, $a \rightarrow 0$} &
		\parbox{13cm}{ \begin{flalign*} \hspace{10pt}
			& \text{\textbf{$C$-metric}} && \nonumber \\[10pt]
			\tensor{\vartheta}{^{\hat{0}}} &= \frac{1}{1 - \alpha r \cos\theta} \sqrt{\left(1 - \alpha^2r^2\right)\left(1 - \frac{2m}{r}\right)} ~ dt && \\
			\tensor{\vartheta}{^{\hat{1}}} &= \frac{1}{1 - \alpha r \cos\theta} \left(\sqrt{\left(1 - \alpha^2r^2\right)\left(1 - \frac{2m}{r}\right)}\right)^{~-1} ~ dr && \\
			\tensor{\vartheta}{^{\hat{2}}} &= \frac{1}{1 - \alpha r \cos\theta} \frac{r}{\sqrt{1 - 2\alpha m \cos\theta}} d\theta && \\
			\tensor{\vartheta}{^{\hat{3}}} &= \frac{1}{1 - \alpha r \cos\theta} \sqrt{1 - 2\alpha m \cos\theta} \, r\sin\theta \, d\phi &&
		\end{flalign*} } \\
	
	\end{tabular}
	\caption{Various coframes. The metric is given by $g = -\tensor{\vartheta}{^{\hat{0}}}\otimes\tensor{\vartheta}{^{\hat{0}}} + \tensor{\vartheta}{^{\hat{1}}}\otimes\tensor{\vartheta}{^{\hat{1}}} + \tensor{\vartheta}{^{\hat{2}}}\otimes\tensor{\vartheta}{^{\hat{2}}} + \tensor{\vartheta}{^{\hat{3}}}\otimes\tensor{\vartheta}{^{\hat{3}}}$.}
	\label{table:coframes}
\end{table}

\section{Curvature}
\label{sec:curvature}
Let us now turn to the curvature of the PD solution. Any antisymmetric 2-form has $\tvect{n}{2}\tvect{n}{2}$ independent components. The curvature components, however, are further constrained by the first Bianchi identity $0 = DD \tensor{\vartheta}{^\mu} = \tensor{\text{Riem}}{^\mu_\alpha} \wedge \tensor{\vartheta}{^\alpha}$. This is a vector-valued 3-form with $n\tvect{n}{3}$ independent components. Therefore, the curvature has $\# = \frac{1}{12}n^2(n+1)(n-1)$ independent components. In four dimensions, $\# = 20$.

We decompose the curvature into its irreducible pieces with respect to the Lorentz group \cite{Garcia:2003bw}:
\begin{align}
\tensor{\text{Riem}}{_\mu_\nu} \quad =: \quad  \tensor{\text{Weyl}}{_\mu_\nu} ~ + ~ \tensor{\cancel{\text{Ricci}}}{_\mu_\nu} ~ + ~ \tensor{\text{Scalar}}{_\mu_\nu} \label{eq:decomposition}
\end{align}
This decomposition consists of three pieces:
\begin{itemize}
\item the (tracefree) Weyl curvature $\tensor{\text{Weyl}}{_\mu_\nu} = \frac{1}{2} \tensor{\text{Weyl}}{_\alpha_\beta_\mu_\nu} \tensor{\vartheta}{^\alpha} \wedge \tensor{\vartheta}{^\beta}$
\item the tracefree Ricci part $\tensor{\cancel{\text{Ricci}}}{_\mu_\nu} := \frac{2}{n-2}\tensor{\vartheta}{_[_\mu} \wedge \tensor{\cancel{\text{Ric}}}{_\nu_]}$ with  $\tensor{\cancel{\text{Ric}}}{_\mu} := \tensor{\text{Ric}}{_\mu} - \frac{1}{n}\text{R}\tensor{\vartheta}{_\mu}$, whereas the Ricci 1-form is given by $\tensor{\text{Ric}}{_\mu} := \tensor{e}{_\alpha} \righthalfcup \tensor{\text{Riem}}{^\alpha_\mu}$.
\item the scalar part $\tensor{\text{Scalar}}{_\mu_\nu} := \frac{1}{n(n-1)} \text{R} \tensor{\vartheta}{_\mu} \wedge \tensor{\vartheta}{_\nu}$ with $\text{R} := \tensor{e}{_\alpha} \righthalfcup \tensor{\text{Ric}}{^\alpha} $
\end{itemize}
Counting degrees of freedom, Eq.\ (\ref{eq:decomposition}) translates into
\begin{align}
20 \text{~(Riemann)} \quad = \quad 10 \text{~(Weyl)} ~+~ 9 \text{~(tracefree Ricci)} ~+~ 1 \text{~(Ricci scalar)} \quad . \label{eq:decomposition_dof}
\end{align} 
By means of the Einstein--Maxwell equations (\ref{eq:einstein_maxwell}), we now notice that only the Weyl piece contains non-trivial information about the PD spacetime (or about any other electro-magneto vacuum spacetime with cosmological constant, for that matter):

Since $\tensor{\Sigma}{_\mu}$ is the energy-momentum 3-form of an electromagnetic field, it is traceless. Taking the trace of the dual of the Einstein equation \eqref{eq:einstein_equation_dual} then implies $\text{R} = 4\Lambda$. This is equivalent to $\tensor{\cancel{\text{Ric}}}{_\mu} = 8\pi\tensor{\Sigma}{_\mu}$. Therefore, only the Weyl part of curvature carries truly non-trivial information.

In order to visualize the 20 independent components of $\tensor{\text{Riem}}{_\mu_\nu}$, we use the symmetry properties of its anholonomic components $\tensor{\text{Riem}}{_\alpha_\beta_\mu_\nu} := \tensor{e}{_\beta} \righthalfcup \left( \tensor{e}{_\alpha} \righthalfcup \tensor{\text{Riem}}{_\mu_\nu} \right)$, namely,
\begin{align}
\tensor{\text{Riem}}{_\alpha_\beta_\mu_\nu} ~=~ -\tensor{\text{Riem}}{_\beta_\alpha_\mu_\nu} ~=~ -\tensor{\text{Riem}}{_\alpha_\beta_\nu_\mu} ~ , \quad \tensor{\text{Riem}}{_\alpha_\beta_\mu_\nu} ~=~ \tensor{\text{Riem}}{_\mu_\nu_\alpha_\beta} ~, \quad  \tensor{\text{Riem}}{_[_\alpha_\beta_\mu_\nu_]} ~ = ~ 0 ~ . \label{eq:riemann_symmetries}
\end{align}
By means of Eq.\ (\ref{eq:decomposition}), this symmetry holds for all pieces of the (irreducible!) decomposition. The symmetries (\ref{eq:riemann_symmetries}) allow us to organize all 20 components in a $6 \times 6$ matrix. It is now convenient to introduce collective anholonomic indices; we define
\begin{alignat}{6}
I, J & \in \{\hat{0}\hat{1}, \hat{0}\hat{2}, \hat{0}\hat{3}, \hat{2}\hat{3}, \hat{3}\hat{1}, \hat{1}\hat{2} \} & ~ \mapsto ~ \{1,2,3,4,5,6\} .
\end{alignat}
The components of the metric \tvect{2}{0} tensor on this six-dimensional space are given by the 0-(pseudo-)form $\tensor{\eta}{^\alpha^\beta^\mu^\nu}$, such that in our conventions (see appendix \ref{appendix:exterior_calculus})
\begin{align}
\Big( \tensor{\eta}{^{\,I}^J} \Big) \quad = \quad \left(
	\begin{array}{cc}
		0 & -\mathbbm{1} \\
		-\mathbbm{1} & 0 \\
	\end{array} \right) , \label{eq:eta_metric}
\end{align}
with $\mathbbm{1} := \text{diag}\left( 1, 1, 1 \right)$. By means of the symmetry \eqref{eq:riemann_symmetries}${}_3$ the trace of this matrix vanishes, $\tensor{\eta}{^{\,A}^B}\,\tensor{\text{Riem}}{_A_B} = 0$, and it comprises in fact 20 independent degrees of freedom. The $6 \times 6$ curvature matrix then reads
\begin{align}
\Big( \tensor{\text{Riem}}{_I_J} \Big) \quad = \quad \left(
	\begin{array}{cccccc}
		-2\mathbb{E} & 0 & 0 & 2\mathbb{B} & 0 & 0 \\
		. & \mathbb{E} & 0 & 0 & -\mathbb{B} & 0 \\
		. & . & \mathbb{E} & 0 & 0 & -\mathbb{B} \\
		. & . & . & 2\mathbb{E} & 0 & 0 \\
		. & . & . & . & -\mathbb{E} & 0 \\
		. & . & . & . & . & -\mathbb{E} \\
	\end{array} \right) \quad + \quad \text{diag}\,\Big(\mathbb{Q},0,0,\mathbb{Q},0,0\Big) \quad + \quad \frac{\Lambda}{3} ~ \check{\mathbbm{1}} \quad . \label{eq:riemann_6x6}
\end{align}
The dots ``.'' denote matrix entries following directly from the symmetry. We defined
\begin{align}
\mathbb{E} := -\frac{1}{2} \tensor{\text{Weyl\,}}{_{\hat{0}}_{\hat{1}}_{\hat{0}}_{\hat{1}}} , \quad \mathbb{B} := \frac{1}{2} \tensor{\text{Weyl\,}}{_{\hat{0}}_{\hat{1}}_{\hat{2}}_{\hat{3}}} , \quad \mathbb{Q} := -2 \star \left( \tensor{\vartheta}{^{\hat{0}}} \wedge \tensor{\Sigma}{_{\hat{0}}} \right) ,
\end{align}
and $\check{\mathbbm{1}} := \text{diag} \, \Big( 1, 1, 1, -1, -1, -1\Big)$. For PD coordinates we find
\begin{alignat}{3}
\mathbb{E} & = {\left(\frac{pq - 1}{p^2 + q^2}\right)}^3 \Big[ (3p^2 - q^2)\hat{m}q + (p^2 - 3q^2)\hat{n}p - (\hat{e}^2 + \hat{g}^2)(p^2 - q^2)(1 + pq) \Big] , \\
\mathbb{B} & = {\left(\frac{pq - 1}{p^2 + q^2}\right)}^3 \Big[ (p^2 - 3q^2)\hat{m}p - (3p^2 - q^2)\hat{n}q + 2(\hat{e}^2 + \hat{g}^2)(1 + pq)pq \Big] , \\
\mathbb{Q} & = \frac{(pq - 1)^4}{(p^2 + q^2)^2} \left( \hat{e}^2 + \hat{g}^2 \right) .
\end{alignat}
As anholonomic components, the above expressions are coordinate independent. Therefore, we can obtain the respective expression in GP coordinates simply by replacing the coordinates and constants according to Eqs.\ \eqref{eq:gp_coordinate_transformation}, \eqref{eq:gp_constant_redefinition}:
\begin{alignat}{4}
\mathbb{E} & = \frac{\Omega^3}{\rho^6} \Big\{ && \left[r^2 - 3(\ell + a\cos\theta)^2\right]mr + \left[3r^2 - (\ell + a\cos\theta)^2\right]n(\ell + a\cos\theta) \nonumber \\
		   &								  && - (e^2 + g^2)\left[r^2 - (\ell + a\cos\theta)^2\right]\left[1 + \alpha r(\ell + a\cos\theta)\right] \Big\} , \label{eq:ee} \\
\mathbb{B} & = \frac{\Omega^3}{\rho^6} \Big\{ && \left[3r^2 - (\ell + a\cos\theta)^2\right]m(\ell + a\cos\theta) - \left[r^2 - 3(\ell + a\cos\theta)^2\right]nr \nonumber \\
		   &								  && - 2(e^2 + g^2)\left[1 + \alpha r(\ell + a\cos\theta)\right]r(\ell + a\cos\theta) \Big\} , \label{eq:bb} \\
\mathbb{Q} & = \frac{\Omega^4}{\rho^4} \Big( && e^2 + g^2 \Big). \label{eq:qq}
\end{alignat}
We recognize the canonical form of curvature for a type D spacetime, as pointed out e.g.\ by Bel \cite{Bel:1962}, Eq.\ (30), and identify Bel's $\alpha$ and $\beta$ with our $\mathbb{E}$ and $-\mathbb{B}$. Following Matte \cite{Matte:1953}, the following tensors completely specify the Weyl curvature:
\begin{align}
\tensor{E}{_\mu_\nu} := \tensor{\text{Weyl}}{_\mu_\alpha_\nu_\beta} \tensor{u}{^\alpha} \tensor{u}{^\beta}, \quad \tensor{B}{_\mu_\nu} := -\tensor{\ast\!\text{Weyl}}{_\mu_\alpha_\nu_\beta} \tensor{u}{^\alpha} \tensor{u}{^\beta} ,
\end{align}
where $u$ is a timelike unit vector, $g(u,u) = -1$, and by the symmetry properties of the Weyl tensor $E$ and $B$ are symmetric, tracefree \tvect{0}{2} tensors. Furthermore, $u$ is an eigenvector to both $\tensor{E}{_\mu_\nu}$ and $\tensor{B}{_\mu_\nu}$ with eigenvalue 0. Therefore, in $n=4$ dimensions, both have $10 - 1 - 4 = 5$ independent components, thereby completely specifying the Weyl curvature. They read
\begin{align}
\left(\tensor{E}{_\mu_\nu}\right) = \mathbb{E} \, \left(
	\begin{array}{cccc}
		0 & 0 & 0 & 0 \\
		. & -2 & 0 & 0 \\
		. & . & 1 & 0 \\
		. & . & . & 1 \\
	\end{array} \right), \quad
\left(\tensor{B}{_\mu_\nu}\right) = \mathbb{B} \, \left(
	\begin{array}{cccc}
		0 & 0 & 0 & 0 \\
		. & -2 & 0 & 0 \\
		. & . & 1 & 0 \\
		. & . & . & 1 \\
	\end{array} \right) .
\end{align}
The dual of the Weyl part of the decomposition of Eq.\ (\ref{eq:riemann_6x6}) turns out to be
\begin{align}
\Big( \tensor{\star \text{Weyl}}{_I_J} \Big) \quad = \quad \left(
	\begin{array}{cccccc}
		2\mathbb{B} & 0 & 0 & 2\mathbb{E} & 0 & 0 \\
		. & -\mathbb{B} & 0 & 0 & -\mathbb{E} & 0 \\
		. & . & -\mathbb{B} & 0 & 0 & -\mathbb{E} \\
		. & . & . & -2\mathbb{B} & 0 & 0 \\
		. & . & . & . & \mathbb{B} & 0 \\
		. & . & . & . & . & \mathbb{B} \\
	\end{array} \right) \quad .
\end{align}
Conversely, the dual swaps $\mathbb{E}$ and $\mathbb{B}$, up to a sign, i.e. $\mathbb{E} \mapsto -\mathbb{B}$ and $\mathbb{B} \mapsto \mathbb{E}$. We adopt a complex null tetrad according to
\begin{align}
l := \frac{1}{\sqrt{2}}\left( \tensor{\vartheta}{^{\hat{0}}} + \tensor{\vartheta}{^{\hat{1}}} \right), \quad  k := \frac{1}{\sqrt{2}}\left( \tensor{\vartheta}{^{\hat{0}}} - \tensor{\vartheta}{^{\hat{1}}} \right), \quad m := \frac{1}{\sqrt{2}}\left( \tensor{\vartheta}{^{\hat{2}}} - i\tensor{\vartheta}{^{\hat{3}}} \right),  \quad \overline{m} := \frac{1}{\sqrt{2}}\left( \tensor{\vartheta}{^{\hat{2}}} + i\tensor{\vartheta}{^{\hat{3}}} \right) , \label{eq:complex_null_tetrad}
\end{align}
such that $\tensor{l}{^\alpha}\tensor{k}{_\alpha} = -1$, $\tensor{m}{^\alpha}\tensor{\overline{m}}{_\alpha} = 1$, with $i^2 = -1$. The only non-vanishing Weyl scalar is
\begin{align}
\Psi_2 := \frac{1}{2} \tensor{\text{Weyl}}{_\alpha_\beta_\gamma_\delta} \tensor{l}{^\alpha} \tensor{k}{^\beta} \left( \tensor{l}{^\gamma}\tensor{k}{^\delta} - \tensor{m}{^\gamma}\tensor{\overline{m}}{^\delta} \right) = - \mathbb{E} - i \, \mathbb{B} ,
\end{align}
representing the ``Coulomb'' component of a spacetime (Szekeres \cite{Szekeres:1965}, Stephani \textit{et al}.\ \cite{Stephani:2003}).

We now take a closer look at $\mathbb{E}$ and $\mathbb{B}$, see Eqs.\ \eqref{eq:ee} and \eqref{eq:bb}. They are exact expressions obtained from direct calculation. For small black hole parameters allowing for asymptotic flatness (small $\ell$, $\Lambda$, and $\alpha$) we have the asymptotic behavior $\Omega \sim 1$ and $\rho \sim r$, and hence
\begin{align}
r^3 \, \mathbb{E} &\sim  m + \frac{3n\left(\ell + a\cos\theta\right)}{r} - \frac{e^2 + g^2}{r} , \\
r^3 \, \mathbb{B} &\sim -n + \frac{3m\left(\ell + a\cos\theta\right)}{r} - \frac{2\left(e^2+g^2\right)\left(\ell + a\cos\theta\right)}{r^2} .
\end{align}
The leading terms are $m$ and $n$, corresponding to gravitoelectric and gravitomagnetic monopoles, respectively. This is why they are sometimes paired up as $m + in$, just like $a + i\ell$ and $e + ig$, see e.g.\ Pleba\'nski and Demia\'nski \cite{Plebanski:1976gy}. Moreover, the gravitoelectric and gravitomagnetic dipoles show up in next-to-leading order, along with the electric and magnetic charges of classical electrodynamics.

For vanishing $n$ and $\ell$ (no gravitomagnetic monopoles), $\mathbb{E}$ and $\mathbb{B}$ can be intuitively interpreted as gravitoelectric charge (mass) and gravitoelectric current (angular momentum). The situation is the same as in classical electrodynamics with no magnetic monopoles. On the other hand, for non-vanishing gravitomagnetic charges the situation remains similar to electrodynamics with magnetic charges. In the limiting Kerr case of the PD solution, we have $\mathbb{E} \propto m$ and $\mathbb{B} \propto ma$.

We conclude: Using the coordinates by Griffiths and Podolsk\'y, $\mathbb{E}$ and $\mathbb{B}$ can be understood in a precise, physical sense, which surpasses a purely formal analogy.

\section{Curvature invariants}
\label{sec:invariants}

The Kretschmann invariant $K$ and the Chern--Pontryagin pseudo-invariant $\mathcal{P}$ can be defined as squares of the Riemannian curvature:
\begin{alignat}{3}
K & := \frac{1}{2} \, \tensor{\text{Riem}}{_\alpha_\beta_\gamma_\delta} \, \tensor{\text{Riem}}{^\alpha^\beta^\gamma^\delta} & ~ = ~ & - \, \star \left[ \tensor{\text{Riem}}{_\alpha_\beta} \wedge \left( \star \tensor{\text{Riem}}{^\alpha^\beta} \right) \right] , \label{eq:kretschmann_definition} \\
\mathcal{P} & := \frac{1}{2} \left( \text{*\,} \tensor{\text{Riem}}{_\alpha_\beta_\gamma_\delta} \right) \, \tensor{\text{Riem}}{^\alpha^\beta^\gamma^\delta} & ~ = ~ & \star \left( \tensor{\text{Riem}}{_\alpha_\beta} \wedge \tensor{\text{Riem}}{^\alpha^\beta} \right) . \label{eq:pontryagin_definition}
\end{alignat}
Here, ``$\star$'' denotes the Hodge dual acting on forms, and ``\,*\,'' denotes the left tensor dual acting on the left pair of antisymmetric indices. We can now use the decomposition (\ref{eq:decomposition}). Since it is irreducible, the individual parts are orthogonal with respect to each other, such that upon squaring there appear no cross terms, see Garc\'ia \textit{et al}.\ \cite{Garcia:2003bw}:
\begin{alignat}{7}
K &=:& K^\text{Weyl} ~&+&~ K^\text{\cancel{Ric}} ~&+&~ K^\text{R} \quad , \\
\mathcal{P} &=:& \mathcal{P}^\text{Weyl} ~&+&~ \mathcal{P}^\text{\cancel{Ric}} ~&+&~ \mathcal{P}^\text{R} \quad .
\end{alignat}
It is well known \cite{Bel:1962} that for any spacetime the Weyl (pseudo-)invariants can be expressed as
\begin{align}
K^\text{Weyl} = \tensor{E}{^\alpha_\beta}\tensor{E}{^\beta_\alpha} - \tensor{B}{^\alpha_\beta}\tensor{B}{^\beta_\alpha} , \quad \mathcal{P}^\text{Weyl} = -2 \tensor{E}{^\alpha_\beta} \tensor{B}{^\beta_\alpha} .
\end{align}
This is a formal similarity to electromagnetism. For the PD solution we obtain
\begin{alignat}{10}
&& K^\text{Weyl} ~ &=&& ~ -24 \left( \mathbb{B}^2 - \mathbb{E}^2 \right) ~ , \quad && K^\text{\cancel{Ric}} ~ &=&& ~ 4 \mathbb{Q}^2 ~ , \quad && K^\text{R} ~ &=&& ~ \frac{4}{3} \Lambda^2 ~ , \label{eq:kretschmann_ed} \\
&& \mathcal{P}^\text{Weyl} ~ &=&& ~ -48 \mathbb{E} \mathbb{B} ~ , \quad && \mathcal{P}^\text{\cancel{Ric}} ~ &=&& ~ 0 ~ , \quad && \mathcal{P}^\text{R} ~ &=&& ~ 0 ~ . \label{eq:pontryagin_ed}
\end{alignat}
For the Kretschmann scalar $K$ of the Kerr spacetime, this result is well-known, see O'Neill \cite{ONeill:1995}, theorem 2.7.2 (for the definition of the two functions $I$ and $J$, here referred to as $\mathbb{E}$ and $\mathbb{B}$, respectively) and corollary 2.7.5 for the form of the Kretschmann scalar (the relative factor 2 arises due to our definition of Kretschmann, see Eq.\ \eqref{eq:kretschmann_definition}).

With the considerations from Sec.\ \ref{sec:curvature}, we see that relations \eqref{eq:kretschmann_ed} and \eqref{eq:pontryagin_ed} also reflect a physical analogy between general relativity and classical electrodynamics with magnetic charges.

\section{Bel and Bel--Robinson tensors, Bel--Robinson 3-form}

With the curvature invariants taking a form so closely related to the invariants of vacuum electrodynamics, we will now try to establish further analogies between the energy momentum of an electric field and its gravitational almost-counterpart, the Bel and Bel--Robinson tensors.

The Bel tensor can be defined via the tensor dual, see Senovilla \cite{Senovilla:1999qv}:
\begin{align}
\begin{split}
2\tensor{B}{_\mu_\nu_\rho_\sigma} := \quad & \tensor{\text{Riem}}{_\mu_\alpha_\beta_\rho} \tensor{\text{Riem}}{_\nu^\alpha^\beta_\sigma} + \left( \text{*\,}\tensor{\text{Riem*\,}}{_\mu_\alpha_\beta_\rho} \right) \left( \text{*\,}\tensor{\text{Riem*\,}}{_\nu^\alpha^\beta_\sigma} \right) \\
									+ & \left( \text{*\,}\tensor{\text{Riem}}{_\mu_\alpha_\beta_\rho} \right) \left( \text{*\,} \tensor{\text{Riem}}{_\nu^\alpha^\beta_\sigma} \right) + \left( \tensor{\text{Riem*\,}}{_\mu_\alpha_\beta_\rho} \right) \left( \tensor{\text{Riem*\,}}{_\nu^\alpha^\beta_\sigma} \right) \label{eq:bel_tensor_definition_duals}
\end{split}
\end{align}
It has the following symmetries:
\begin{align}
\tensor{B}{_[_\mu_\nu_]_\rho_\sigma} = \tensor{B}{_\mu_\nu_[_\rho_\sigma_]} = 0, \quad \tensor{B}{_\mu_\nu_\rho_\sigma} = \tensor{B}{_\rho_\sigma_\mu_\nu}, \quad \tensor{B}{^\alpha_\alpha_\rho_\sigma} = 0
\end{align}
Note that $\tensor{B}{^\alpha_\mu_\alpha_\sigma} \neq 0$. The Bel--Robinson tensor can be defined as (Senovilla \cite{Senovilla:1999qv})
\begin{align}
\tensor{\widetilde{B}}{_\mu_\nu_\rho_\sigma} := \tensor{\text{Weyl}}{_\mu_\alpha_\beta_\rho} \tensor{\text{Weyl}}{_\nu^\alpha^\beta_\sigma} + \left( \text{*\,}\tensor{\text{Weyl}}{_\mu_\alpha_\beta_\rho} \right) \left( \text{*\,}\tensor{\text{Weyl}}{_\nu^\alpha^\beta_\sigma} \right) \label{eq:bel_robinson_definition} .
\end{align}
It is completely symmetric in all its indices and completely tracefree \cite{Robinson:1997,Senovilla:1999qv,So:2010yq}:
\begin{align}
\tensor{\widetilde{B}}{_\mu_\nu_\rho_\sigma} = \tensor{\widetilde{B}}{_(_\mu_\nu_\rho_\sigma_)} , \quad \tensor{\widetilde{B}}{^\alpha_\nu_\alpha_\sigma} = 0 \label{eq:bel_robinson_symmetries}
\end{align}

There are several other definitions in the literature for the Bel and Bel--Robinson tensors, for a review and comparison of definitions in four dimensions see Douglas \cite{Douglas:2003}, for an extensive review, valid in all spacetime dimensions, see Senovilla \cite{Senovilla:1999xz}. In the following we will use the definitions for the Bel and Bel--Robinson tensors given above.

The Bel and Bel--Robinson tensors are interesting, since they are the closest tensorial objects available to describe gravitational energy momentum (see e.g. Garecki \cite{Garecki:1985, Garecki:2000dj}, Mashhoon \cite{Mashhoon:1998tt}, and the references above). However, Eqs.\ \eqref{eq:bel_tensor_definition_duals} and \eqref{eq:bel_robinson_definition} do not allow such a conclusion yet. Therefore we will motivate this interpretation briefly by employing analogies from electrodynamics:

Expressed in terms of components $F = \frac{1}{2} \tensor{F}{_\alpha_\beta} \tensor{\vartheta}{^\alpha} \wedge \tensor{\vartheta}{^\beta}$, the symmetric tracefree electromagnetic energy momentum \tvect{0}{2} tensor defined via $\tensor{T}{_\mu_\nu} := \tensor{e}{_\mu} \righthalfcup \star \tensor{\Sigma}{_\nu}$ can be written as
\begin{align}
\tensor{T}{_\mu_\nu} = \frac{1}{2} \left[ \tensor{F}{_\mu_\alpha} \tensor{F}{^\alpha_\nu} + \left( \tensor{\ast F}{_\mu_\alpha} \right) \left( \tensor{\ast F}{^\alpha_\nu} \right) \right] . \label{eq:electromagnetic_energy_momentum_duals}
\end{align}
This form is quite similar to Eq.\ \eqref{eq:bel_tensor_definition_duals}. Furthermore, inserting the Riemannian curvature 2-form into Eq.\ \eqref{eq:electrodynamics_energy_momentum_3_form} and contracting over both indices yields
\begin{align}
\begin{split}
\tensor{B}{_\mu} &:= \frac{1}{2} \left[ \tensor{\text{Riem}}{_\alpha_\beta} \wedge \left( \tensor{e}{_\mu} \righthalfcup \star \tensor{\text{Riem}}{^\alpha^\beta} \right) - \left( \star \tensor{\text{Riem}}{_\alpha_\beta} \right) \wedge \left( \tensor{e}{_\mu} \righthalfcup \tensor{\text{Riem}}{^\alpha^\beta} \right) \right] \\
				 &= \frac{1}{4} \tensor{\text{Riem}}{_\alpha_\beta_\gamma_\delta} \tensor{\text{Riem}}{^\alpha^\beta^\gamma^\delta} \tensor{\eta}{_\mu} - \tensor{\text{Riem}}{_\mu_\beta_\gamma_\delta} \tensor{\text{Riem}}{^\alpha^\beta^\gamma^\delta} \tensor{\eta}{_\alpha} \label{eq:bel_3_form_no_indices} .
\end{split}
\end{align}
On the other hand, the electromagnetic energy momentum 3-form turns out to be
\begin{align}
\tensor{\Sigma}{_\mu} = \frac{1}{4} \tensor{F}{_\alpha_\beta} \tensor{F}{^\alpha^\beta} \tensor{\eta}{_\mu} - \tensor{F}{_\mu_\alpha} \tensor{F}{^\beta^\alpha} \tensor{\eta}{_\beta} . \label{eq:electromagnetic_energy_momentum_3_form}
\end{align}
The similarity between Eqs.\ \eqref{eq:bel_3_form_no_indices} and \eqref{eq:electromagnetic_energy_momentum_3_form} is obvious. Is it also possible to find a 3-form $\tensor{\Sigma}{_\nu_\rho_\sigma}$, such that $\tensor{B}{_\mu_\nu_\rho_\sigma} = \tensor{e}{_\mu} \righthalfcup \star \tensor{\Sigma}{_\nu_\rho_\sigma}$? For the Bel--Robinson tensor the answer is affirmative:
\begin{align}
\tensor{\widetilde{B}}{_\mu_\nu_\rho_\sigma} = \tensor{e}{_\mu} \righthalfcup \star \, \Big[ \tensor{\text{Weyl}}{_\rho_\alpha} \wedge \left( \tensor{e}{_\nu} \righthalfcup \star \tensor{\text{Weyl}}{^\alpha_\sigma} \right) - \left( \star \tensor{\text{Weyl}}{_\rho_\alpha} \right) \wedge \left( \tensor{e}{_\nu} \righthalfcup \tensor{\text{Weyl}}{^\alpha_\sigma} \right) \Big] =: \tensor{e}{_\mu} \righthalfcup \star \tensor{\widetilde{\Sigma}}{_\nu_\rho_\sigma} . \label{eq:bel_robinson_3_form_correspondence}
\end{align}
We call $\tensor{\widetilde{\Sigma}}{_\nu_\rho_\sigma}$ the the Bel--Robsinson 3-form. Its similarity to the electromagnetic (vacuum) energy-momentum 3-form $\tensor{\Sigma}{_\mu}$ (see Eq.\ \eqref{eq:electrodynamics_energy_momentum_3_form} with $H = \star F$) becomes now fully apparent:
\begin{align}
\tensor{\widetilde{\Sigma}}{_\nu_\rho_\sigma} &:= \tensor{\text{Weyl}}{_\rho_\alpha} \wedge \left( \tensor{e}{_\nu} \righthalfcup \star \tensor{\text{Weyl}}{^\alpha_\sigma} \right) - \left( \star \tensor{\text{Weyl}}{_\rho_\alpha} \right) \wedge \left( \tensor{e}{_\nu} \righthalfcup \tensor{\text{Weyl}}{^\alpha_\sigma} \right) , \label{eq:bel_robinson_3_form} \\
\tensor{\Sigma}{_\mu} &:= \frac{1}{2} \left[ F \wedge \left( \tensor{e}{_\mu} \righthalfcup \star F \right) - \left( \star F \right) \wedge \left( \tensor{e}{_\mu} \righthalfcup F \right) \right] .
\end{align}
Moreover, the Bel--Robinson 3-form is traceless, just like the electromagnetic energy-momentum:
\begin{align}
\tensor{\vartheta}{^\alpha} \wedge \tensor{\widetilde{\Sigma}}{_\alpha_\rho_\sigma} &= 0 \label{eq:bel_robinson_3_form_tracefree} \\
\tensor{\vartheta}{^\alpha} \wedge \tensor{\Sigma}{_\alpha} &= 0
\end{align}
We summarize: The 3-form $\tensor{\tilde{\Sigma}}{_\nu_\rho_\sigma}$ is (up to a factor of 2) the energy momentum of Eq.\ \eqref{eq:electrodynamics_energy_momentum_3_form} where we replaced the 2-form $F$ with the Weyl curvature 2-form. The only modification arises due to the tensorial indices of the Riemann curvature 2-form. After performing the only possible non-trivial trace (summation over $\alpha$, unique up to a sign) we end up with the correct energy momentum \tvect{0}{3}-valued 3-form. See the proof of Eqs.\ \eqref{eq:bel_robinson_3_form_correspondence} and \eqref{eq:bel_robinson_3_form_tracefree} in appendix \ref{appendix:bel_robinson_3_form}.

The Bel--Robinson 3-form has been found recently in an interesting paper by G\'omez-Lobo \cite{Gomez-Lobo:2014xsa}. In the present work, see Eq.\ \eqref{eq:bel_robinson_3_form}, we used a physically motivated approach to arrive at the Bel--Robinson 3-form very quickly. We also show in appendix \ref{appendix:bel_robinson_3_form} that the two formulations are in fact equivalent.

For the Bel tensor this procedure is not straightforward, because the symmetry $\tensor{B}{_\mu_\nu_\rho_\sigma} = \tensor{B}{_\rho_\sigma_\mu_\nu}$ has to be put in by hand, and a part of its trace has to be subtracted as well:
\begin{align}
\tensor{B}{_\mu_\nu_\rho_\sigma} & = \tensor{e}{_\mu} \righthalfcup \star \tensor{\Sigma}{_\nu_\rho_\sigma} + \tensor{e}{_\rho} \righthalfcup \star \tensor{\Sigma}{_\sigma_\mu_\nu} - \frac{1}{2} \left( \tensor{g}{_\mu_\nu} \tensor{\text{Tr}}{_\rho_\sigma} + \tensor{g}{_\rho_\sigma} \tensor{\text{Tr}}{_\mu_\nu} \right) + \frac{1}{8} \tensor{g}{_\mu_\nu}\tensor{g}{_\rho_\sigma} \tensor{\text{Tr}}{^\alpha_\alpha} , \\
\tensor{\Sigma}{_\nu_\rho_\sigma} &:= \tensor{\text{Riem}}{_\rho_\alpha} \wedge \left( \tensor{e}{_\nu} \righthalfcup \star \tensor{\text{Riem}}{^\alpha_\sigma} \right) - \left( \star \tensor{\text{Riem}}{_\rho_\alpha} \right) \wedge \left( \tensor{e}{_\nu} \righthalfcup \tensor{\text{Riem}}{^\alpha_\sigma} \right) , \\
\tensor{\text{Tr}}{_\mu_\nu} & := \frac{1}{2} \left( \tensor{e}{_\mu} \righthalfcup \star \tensor{\Sigma}{_\nu^\alpha_\alpha} + \tensor{e}{_\alpha} \righthalfcup \star \tensor{\Sigma}{^\alpha_\mu_\nu} \right) .
\end{align}
These complications are rooted in the following property of the Weyl tensor, that is not valid for the Riemann tensor (only in vacuum, where they coincide) \cite{Edgar:2001vv,Wingbrant:2003}:
\begin{align}
\tensor{\text{Weyl}}{_\mu_\alpha_\beta_\gamma} \tensor{\text{Weyl}}{_\nu^\alpha^\beta^\gamma} = \frac{1}{4} \tensor{g}{_\mu_\nu} \tensor{\text{Weyl}}{_\alpha_\beta_\gamma_\delta} \tensor{\text{Weyl}}{^\alpha^\beta^\gamma^\delta} \label{eq:weyl_tensor_vacuum_relation}
\end{align}
This vacuum relation is essential for the tracelessness of the Bel-Robinson 3-form. Incidentally, see the references above, it is also inherited by the Bel--Robinson tensor. A similar relation holds for the electromagnetic energy momentum, that is, $\tensor{T}{_\mu_\alpha}\tensor{T}{_\nu^\alpha} = \frac{1}{4}\tensor{g}{_\mu_\nu}\tensor{T}{_\alpha_\beta}\tensor{T}{^\alpha^\beta}$.

Therefore, the Bel--Robinson tensor seems to be of greater physical interest in non-vacuum spacetimes. Furthermore, it seems to be the direct analogon of the energy momentum tensor of the electromagnetic field. We introduce collective anholonomic indices
\begin{alignat}{6}
I, J & \in \{\hat{0}\hat{0}, \hat{0}\hat{1}, \hat{0}\hat{2}, \hat{0}\hat{3}, \hat{1}\hat{1}, \hat{1}\hat{2}, \hat{1}\hat{3}, \hat{2}\hat{2}, \hat{2}\hat{3}, \hat{3}\hat{3} \} & ~ \mapsto ~ \{1,2,3,4,5,6,7,8,9,10\}
\end{alignat}
and find for the collective components of the Bel--Robinson tensor for a generic type D spacetime
\begin{align}
\Big( \tensor{\widetilde{B}}{_I_J} \Big) \quad &= \quad \left( \mathbb{E}^2 + \mathbb{B}^2 \right) \left(
	\begin{array}{cccccccccc}
		6 &  0 &  0 &  0 & -2 &  0 &  0 &  4 &  0 &  4 \\
		. & -2 &  0 &  0 &  0 &  0 &  0 &  0 &  0 &  0 \\
		. &  . &  4 &  0 &  0 &  0 &  0 &  0 &  0 &  0 \\
		. &  . &  . &  4 &  0 &  0 &  0 &  0 &  0 &  0 \\
		. &  . &  . &  . &  6 &  0 &  0 & -4 &  0 & -4 \\
		. &  . &  . &  . &  . & -4 &  0 &  0 &  0 &  0 \\
		. &  . &  . &  . &  . &  . & -4 &  0 &  0 &  0 \\
		. &  . &  . &  . &  . &  . &  . &  6 &  0 &  2 \\
		. &  . &  . &  . &  . &  . &  . &  . &  2 &  0 \\
		. &  . &  . &  . &  . &  . &  . &  . &  . &  6 \\
	\end{array} \right) .
\end{align}
The quantity $\mathbb{E}^2 + \mathbb{B}^2$ nicely resembles the (positive definite) vacuum energy density of an electromagnetic field (see e.g. Hehl and Obukhov \cite{Hehl:2003}, Eq.\ (E.1.34)). Note, however, that its physical unit is energy density squared:
\begin{align}
\left[ \mathbb{E}^2 \right] = \left[ \mathbb{B}^2 \right] =  {\left(\frac{\text{energy}}{\text{3-volume}}\right)}^2 .
\end{align}
Algebraically, $\mathbb{E}^2 + \mathbb{B}^2$ turns out to be the the magnitude squared of the Weyl scalar, $\Psi_2\overline{\Psi}_2 = \mathbb{E}^2 + \mathbb{B}^2$. On the other hand, for Petrov type D spacetimes, the Newman--Penrose formalism relates the Kretschmann and Pontryagin (pseudo-)invariants of the Weyl tensor to the Weyl scalar $\Psi_2$ as follows:
\begin{align}
K^\text{Weyl} - i \, \mathcal{P}{^\text{Weyl}} = 24 \left(\Psi_2\right)^2 \quad \Rightarrow \quad 24 \Psi_2 \overline{\Psi}_2 = \sqrt{ {\left(K^\text{Weyl}\right)}^2 + {\left(\mathcal{P}{^\text{Weyl}}\right)}^2 } .
\end{align}
Accordingly, the ``energy density'' quantity should be expressable as an invariant, since the square of the Pontryagin pseudo-invariant is again an invariant. In our pseudo-orthogonal coframe, we find for the PD solution
\begin{align}
24 \left( \mathbb{E}^2 + \mathbb{B}^2 \right) = \sqrt{ \tensor{\widetilde{B}}{_\alpha_\beta_\gamma_\delta} \tensor{\widetilde{B}}{^\alpha^\beta^\gamma^\delta} } . \label{eq:energy_density_as_square_of_bel_robinson}
\end{align}
The analogon of Eq.\ \eqref{eq:energy_density_as_square_of_bel_robinson} within vacuum electrodynamics for a field configuration $(\mathbf{E}, \mathbf{B})$ is
\begin{align}
\sqrt{\tensor{T}{_\alpha_\beta} \tensor{T}{^\alpha^\beta}} = \mathbf{E}^2 + \mathbf{B}^2 ,
\end{align}
evaluated in an inertial frame. Bonilla and Senovilla \cite{Bonilla:1997} interpret the Bel--Robinson tensor as an energy-squared expression. By expanding it in terms of the complex null tetrad $\{l,k,m,\overline{m}\}$, see Eq.\ \eqref{eq:complex_null_tetrad}, they define an effective square root for a completely symmetric, tracefree rank \tvect{0}{4} tensor. According to their Eq.~(16), the symmetric, tracefree square root $\tensor{t}{_\alpha_\beta}$ then reads (for any type D spacetime)
\begin{align}
\tensor{t}{_\alpha_\beta} = \epsilon \, 6 \sqrt{\Psi_2 \overline{\Psi}_2} \left( \tensor{m}{_(_\alpha}\tensor{\overline{m}}{_\beta_)} + \tensor{l}{_(_\alpha} \tensor{k}{_\beta_)} \right) .
\end{align}
$\epsilon = \pm 1$ can be chosen freely. For the PD solution we find:
\begin{align}
\left(\tensor{t}{_\alpha_\beta}\right) = \epsilon \, 3 \, \sqrt{\mathbb{E}^2 + \mathbb{B}^2} \, \text{diag}\left( 1, -1, 1, 1 \right)
\end{align}
This is clearly symmetric and tracefree.

\section{Kummer--Weyl tensor}
With the quadratic expressions given in such a concise form, we may proceed to cubic quantities (defined in terms of $\mathbb{E}$ and $\mathbb{B}$, the following results hold for any type D spacetime). A candidate is the Kummer tensor:
\begin{align}
\tensor{K}{^\mu^\nu^\rho^\sigma}[T] := \tensor{T}{^\alpha^\mu^\beta^\nu} \, \tensor{\text{*}T\text{*}}{_\alpha_\gamma_\beta_\delta} \, \tensor{T}{^\gamma^\rho^\delta^\sigma} . \label{eq:kummer_tensor_definition}
\end{align}
The Kummer tensor can be defined for any tensor $T$ of rank \tvect{0}{4} which is antisymmetric according to $\tensor{T}{_(_\mu_\nu_)_\alpha_\beta} = \tensor{T}{_\mu_\nu_(_\alpha_\beta_)} = 0$. $\tensor{\text{*}T\text{*}}{_\alpha_\beta_\gamma_\delta}$ denotes the double tensor dual. Without taking into account further symmetries that $T$ might have, the Kummer tensor satisfies
\begin{align}
\tensor{K}{^\alpha^\beta^\mu^\nu} = \tensor{K}{^\mu^\nu^\alpha^\beta} . \label{eq:kummer_symmetry}
\end{align}
Therefore, in $n = 4$ dimensions, the Kummer tensor can be thought of as a symmetric $16 \times 16$ matrix with 136 independent components. See the recent article by Baekler \textit{et al}.\ \cite{Baekler:2014kha} for an extensive and systematic introduction of the Kummer tensor.

They decompose the Kummer tensor into six pieces ${}^{(I)}K$, with $I = 1, \dots, 6$. In terms of degrees of freedom, $136 = 35 + 45+ 20 + 20 + 15 + 1$. The pieces read (see \cite{Baekler:2014kha}, Eqs.\ (90)--(94), (99), and (100)):
\begin{align}
\begin{split}
{}^{(1)}K{}^{\alpha\beta\mu\nu} & := K{}^{(\alpha\beta\mu\nu)} , \\
{}^{(2)}K{}^{\alpha\beta\mu\nu} & := \frac{1}{2} \left( K{}^{(\alpha|\beta|\mu)\nu} - K{}^{(\beta|\alpha|\nu)\mu} \right) , \\
{}^{(3)}K{}^{\alpha\beta\mu\nu} & := \frac{1}{3} \left( K{}^{\alpha\beta(\mu\nu)} - K{}^{\alpha(\nu\mu)\beta} + K{}^{\beta\alpha(\mu\nu)} - K{}^{\beta(\nu\mu)\alpha} \right) , \\
{}^{(4)}K{}^{\alpha\beta\mu\nu} & := \frac{1}{3} \left( K{}^{\alpha\beta[\mu\nu]} + K{}^{\mu\beta[\alpha\nu]} + K{}^{\beta\alpha[\nu\mu]} + K{}^{\beta\mu[\nu\alpha]} \right) , \\
{}^{(5)}K{}^{\alpha\beta\mu\nu} & := \frac{1}{2} \left( K{}^{[\alpha|\beta|\mu]\nu} - K{}^{[\beta|\alpha|\nu]\mu} \right) , \\
{}^{(6)}K{}^{\alpha\beta\mu\nu} & := K{}^{[\alpha\beta\mu\nu]} . \label{eq:kummer_irreducible_decomposition}
\end{split}
\end{align}
It is useful to introduce the two cubic invariants
\begin{align}
S & := \tensor{{}^{(1)}K}{^\alpha_\alpha^\beta_\beta} , \\
\mathcal{A} & := \tensor{\eta}{_\alpha_\beta_\gamma_\delta} \tensor{{}^{(6)}K}{^\alpha^\beta^\gamma^\delta} .
\end{align}
$S$ may be called the Kummer scalar, and $\mathcal{A}$ the (axial) Kummer pseudo-scalar.

We now turn back to general relativity: The Riemann curvature tensor fulfills the required symmetries, see Eq.\ ~\eqref{eq:riemann_symmetries}, and so does the Weyl tensor. Due to the pair commutation symmetry of Riemann and Weyl, both Kummer--Riemann and Kummer--Weyl fulfill the additional symmetry
\begin{align}
\tensor{K}{_\mu_\nu_\alpha_\beta}[\text{Weyl / Riem}] = \tensor{K}{_\nu_\mu_\beta_\alpha}[\text{Weyl / Riem}] .
\end{align}
As shown in Sec.\ ~\ref{sec:curvature}, the Weyl part is the only non-trivial vacuum contribution to curvature. Therefore, in the following we will evaluate the Kummer--Weyl tensor $K[\text{Weyl}]$. This is physically well-motivated: The form of the Kummer--Weyl tensor is related to the principal null directions of curvature, see Baekler \textit{et al}.\ \cite{Baekler:2014kha}.

The irreducible parts can be represented as matrices. ${}^{(1)}K$ is completely symmetric, and can therefore be --- somewhat redundantly --- visualized as a symmetric $10 \times 10$ matrix. For the PD solution, ${}^{(2)}K = {}^{(5)}K = 0$. ${}^{(3)}K$ is symmetric in its first two indices (and by Eq.\ \eqref{eq:kummer_symmetry} also in its second two), that is, $\tensor{{}^{(3)}K}{_\mu_\nu_\alpha_\beta} = \tensor{{}^{(3)}K}{_\nu_\mu_\alpha_\beta}$. This also allows for a $10 \times 10$ representation. ${}^{(4)}K$ does not exhibit any obvious symmetry, therefore it has to be represented as a $16 \times 16$ matrix. Finally, ${}^{(6)}K$ is completely antisymmetric and must therefore be proportional to the $\eta$ metric of Eq.\ \eqref{eq:eta_metric}. We define the following abbreviations:
\begin{alignat}{5}
\mathbb{P}_0 & := -9\mathbb{B}^2 - 5\mathbb{E}^2 , \qquad && \mathbb{P}_1 && := \frac{8}{3} \mathbb{E} \left( -3 \mathbb{B}^2 + 7 \mathbb{E}^2 \right) , \\
\mathbb{P}_2 & := \frac{2}{3} \mathbb{E} \left( 15 \mathbb{B}^2 - 17 \mathbb{E}^2 \right) , \qquad && \mathbb{P}_3 && := 3 \mathbb{B} \left( 5 \mathbb{B}^2 - 3 \mathbb{E}^2 \right) , \\
\mathbb{P}_4 & := \frac{8}{3} \mathbb{E} \left(3 \mathbb{B}^2 - \mathbb{E}^2 \right) , \qquad && \mathbb{P}_5 && := \left( \mathbb{P}_4 \right)^{-1} \mathbb{P}_6 , \\
\mathbb{P}_6 & := 3 \mathbb{B} \left( \mathbb{B}^2 - 3 \mathbb{E}^2 \right) .
\end{alignat}
The completely symmetric and antisymmetric pieces turn out to be
\begin{singlespace}
\begin{align}
\Big( \tensor{{}^{(1)}K[\text{W}]}{_I_J} \Big) ~ &= ~ \mathbb{E} \, \left(
	\begin{array}{cccccccccc}
		-12\mathbb{E}^2 & 0 & 0 & 0 & 4\mathbb{E}^2 & 0 & 0 & \mathbb{P}_0 & 0 & \mathbb{P}_0 \\
		. & 4\mathbb{E}^2 & 0 & 0 & 0 & 0 & 0 & 0 & 0 & 0 \\
		. & . & \mathbb{P}_0 & 0 & 0 & 0 & 0 & 0 & 0 & 0 \\
		. & . & . & \mathbb{P}_0 & 0 & 0 & 0 & 0 & 0 & 0 \\
		. & . & . & . & -12\mathbb{E}^2 & 0 & 0 & -\mathbb{P}_0 & 0 & -\mathbb{P}_0 \\
		. & . & . & . & . & -\mathbb{P}_0 & 0 & 0 & 0 & 0 \\
		. & . & . & . & . & . & -\mathbb{P}_0 & 0 & 0 & 0 \\
		. & . & . & . & . & . & . & -12\mathbb{E}^2 & 0 & -4\mathbb{E}^2 \\
		. & . & . & . & . & . & . & . & -4\mathbb{E}^2 & 0 \\
		. & . & . & . & . & . & . & . & . & -12\mathbb{E}^2
	\end{array} \right) , \\[15pt]
\Big( \tensor{{}^{(6)}K[\text{W}]}{_I_J} \Big) ~ &= ~ \frac{1}{3} \, \mathbb{P}_6 \, \left( \tensor{\eta}{_I_J} \right) .
\end{align}
\end{singlespace}
The invariants read
\begin{align}
S[\text{Weyl}] & = 24 \mathbb{E}\left( 3\mathbb{B}^2 - \mathbb{E}^2 \right) , \\
\mathcal{A}[\text{Weyl}] & = 24 \mathbb{B} \left( 3 \mathbb{E}^2 - \mathbb{B}^2 \right) .
\end{align}
We recognize Bel's ``fundamental vacuum scalars'', $D$ and $E$, cubic in curvature, see Eqs.\ (11) and (15) in Bel \cite{Bel:1962}. They correspond to the two cubic Weyl invariants in four dimensions, see Fulling \textit{et al}.\ \cite{Fulling:1992vm}, p.\ 1158.

All other components of Kummer may be written in terms of simple expressions $\mathbb{E}\left( \alpha \mathbb{B}^2 - \beta \mathbb{E}^2 \right)$ or $\mathbb{B}\left( \gamma \mathbb{B}^2 - \delta \mathbb{E}^2 \right)$. The symmetric part ${}^{(1)}K$ is proportional to the electric part $\mathbb{E}$, whereas the antisymmetric part ${}^{(6)}K$ is proportional to the magnetic part $\mathbb{B}$. The same holds for their invariants $S$ and $\mathcal{A}$. The results agree with the respective expressions for the \emph{Kerr metric}, first obtained by Baekler \cite{Baekler:2014b}.

The ${}^{(3)}K$ piece is neither proportional to $\mathbb{E}$ or $\mathbb{B}$:
\begin{singlespace}
\begin{align}
\Big( \tensor{{}^{(3)}K[\text{W}]}{_I_J} \Big) ~ = ~ \left(
	\begin{array}{cccccccccc}
		0 & 0 & 0 & 0 & \mathbb{P}_1 & 0 & 0 & \mathbb{P}_2 & 0 & \mathbb{P}_2 \\
		. & -\frac{1}{2}\mathbb{P}_1 & 0 & 0 & 0 & 0 & 0 & 0 & 0 & 0 \\
		. & . & -\frac{1}{2}\mathbb{P}_2 & 0 & 0 & 0 & \mathbb{P}_3 & 0 & 0 & 0 \\
		. & . & . & -\frac{1}{2}\mathbb{P}_2 & 0 & -\mathbb{P}_3 & 0 & 0 & 0 & 0 \\
		. & . & . & . & 0 & 0 & 0 & -\mathbb{P}_2 & 0 & -\mathbb{P}_2 \\
		. & . & . & . & . & \frac{1}{2}\mathbb{P}_2 & 0 & 0 & 0 & 0 \\
		. & . & . & . & . & . & \frac{1}{2}\mathbb{P}_2 & 0 & 0 & 0 \\
		. & . & . & . & . & . & . & 0 & 0 & -\mathbb{P}_1 \\
		. & . & . & . & . & . & . & . & \frac{1}{2}\mathbb{P}_1 & 0 \\
		. & . & . & . & . & . & . & . & . & 0
	\end{array} \right)
\end{align}
\end{singlespace}
The piece ${}^{(4)}K$ reads
\begin{singlespace}
\begin{align}
\Big( \tensor{{}^{(4)}K[\text{W}]}{_I_J} \Big) ~ &= ~ \mathbb{P}_4 \left(
	\begin{array}{cccccccccccccccc}
0 & 0 & 0 & 0 & 0 & 1 & 0 & 0 & 0 & 0 & -\frac{1}{8} & 0 & 0 & 0 & 0 & -\frac{1}{8} \\
. & -2 & 0 & 0 & 1 & 0 & 0 & 0 & 0 & 0 & 0 & \mathbb{P}_5 & 0 & 0 & -\mathbb{P}_5 & 0 \\
. & . & \frac{1}{4} & 0 & 0 & 0 & 0 & 0 & -\frac{1}{8} & 0 & 0 & 0 & 0 & -\mathbb{P}_5 & 0 & 0 \\
. & . & . & \frac{1}{4} & 0 & 0 & 0 & 0 & 0 & \mathbb{P}_5 & 0 & 0 & -\frac{1}{8} & 0 & 0 & 0 \\
. & . & . & . & -2 & 0 & 0 & 0 & 0 & 0 & 0 & -\mathbb{P}_5 & 0 & 0 & \mathbb{P}_5 & 0 \\
. & . & . & . & . & 0 & 0 & 0 & 0 & 0 & \frac{1}{8} & 0 & 0 & 0 & 0 & \frac{1}{8} \\
. & . & . & . & . & . & -\frac{1}{4} & 0 & 0 & \frac{1}{8} & 0 & 0 & \mathbb{P}_5 & 0 & 0 & 0 \\
. & . & . & . & . & . & . & -\frac{1}{4} & -\mathbb{P}_5 & 0 & 0 & 0 & 0 & \frac{1}{8} & 0 & 0 \\
. & . & . & . & . & . & . & . & \frac{1}{4} & 0 & 0 & 0 & 0 & 0 & 0 & 0 \\
. & . & . & . & . & . & . & . & . & -\frac{1}{4} & 0 & 0 & 0 & 0 & 0 & 0 \\
. & . & . & . & . & . & . & . & . & . & 0 & 0 & 0 & 0 & 0 & -1 \\
. & . & . & . & . & . & . & . & . & . & . & 2 & 0 & 0 & -1 & 0 \\
. & . & . & . & . & . & . & . & . & . & . & . & \frac{1}{4} & 0 & 0 & 0 \\
. & . & . & . & . & . & . & . & . & . & . & . & . & -\frac{1}{4} & 0 & 0 \\
. & . & . & . & . & . & . & . & . & . & . & . & . & . & 2 & 0 \\
. & . & . & . & . & . & . & . & . & . & . & . & . & . & . & 0 \\
	\end{array} \right) .
\end{align}
\end{singlespace}
The irreducible decomposition \eqref{eq:kummer_irreducible_decomposition} holds for any tensor $T$ fed into the Kummer machine \eqref{eq:kummer_tensor_definition}. It is expected, however, that this irreducible decomposition will simplify when the Kummer machine is applied to a tensor of higher symmetry than $T$, say, the Weyl tensor.

\section{Conclusions}
In this paper, we have used the Pleba\'nski--Demia\'nski solution of general relativity to show that the formal analogy of invariants in general relativity and electrodynamics can be understood in a physical sense: the electric field $\mathbf{E}$ (charges) corresponds to the mass parameter $m$ (gravitational charge), and the magnetic field $\mathbf{B}$ (moving charges) corresponds to the angular momentum $ma$ (moving gravitational charges) for vanishing gravitomagnetic monopoles.

Secondly, we rederived the Bel--Robinson 3-form in a novel way which uses the definition of electromagnetic energy-momentum. This derivation is physically straightforward 

Finally, we calculated the irreducible pieces of the Kummer tensor for any type D spacetime. These results are hoped to be helpful for determining the relation of the properties of the Kummer tensor with the principal null directions of curvature for a generic type D spacetime.

\section*{Acknowledgments}

The author is grateful to Friedrich W.\ Hehl (Cologne) for bringing this project under way, as well as for many insightful discussions. The author would also like to thank Christian Heinicke (Cologne) for useful remarks regarding computer algebra, and Claus Kiefer (Cologne) for the motivation and continuous support to publish the findings. Further gratitude is expressed to Jos\'e M.\ M.\ Senovilla (Bilbao) for his comments on an earlier draft of this paper, as well as to Jiri Podolsk\'y (Prague) for helpful remarks. This work was partly supported by a scholarship of the Bonn--Cologne Graduate School of Physics and Astronomy (BCGS).

\appendix

\renewcommand{\theequation}{\Alph{section}.\arabic{equation}}
\setcounter{equation}{0} 

\section{Bel--Robinson 3-form}
\label{appendix:bel_robinson_3_form}

The Bel--Robinson 3-form, see Eq.\ \eqref{eq:bel_robinson_3_form}, is given by
\begin{align}
\tensor{\tilde{\Sigma}}{_\nu_\rho_\sigma} := \tensor{\text{Weyl}}{_\rho_\alpha} \wedge \left( \tensor{e}{_\nu} \righthalfcup \star \tensor{\text{Weyl}}{^\alpha_\sigma} \right) - \left( \star \tensor{\text{Weyl}}{_\rho_\alpha} \right) \wedge \left( \tensor{e}{_\nu} \righthalfcup \tensor{\text{Weyl}}{^\alpha_\sigma} \right) .
\end{align}
We expand the Weyl 2-forms in components, $\tensor{\text{Weyl}}{_\mu_\nu} =: \frac{1}{2} \tensor{\text{Weyl}}{_\alpha_\beta_\mu_\nu} \tensor{\vartheta}{^\alpha} \wedge \tensor{\vartheta}{^\beta}$ and find
\begin{align}
\tensor{\tilde{\Sigma}}{_\nu_\rho_\sigma} = \frac{1}{2} \tensor{\text{Weyl}}{_\omega_\tau_\rho_\alpha} \tensor{\text{Weyl}}{^\omega^\tau^\alpha_\sigma} \tensor{\eta}{_\nu} - \left( \tensor{\text{Weyl}}{_\nu_\tau_\rho_\alpha}\tensor{\text{Weyl}}{^\omega^\tau^\alpha_\sigma} + \tensor{\text{Weyl}}{_\nu_\tau_\sigma_\alpha}\tensor{\text{Weyl}}{^\omega^\tau^\alpha_\rho} \right) \tensor{\eta}{_\omega} . \label{eq:bel_robinson_3_form_intermediate}
\end{align}
We now evaluate the dual of this expression. The Hodge star only acts on the $(n-1)$-forms $\tensor{\eta}{_\mu}$ according to $\star \tensor{\eta}{_\mu} \equiv \star \star \tensor{\vartheta}{_\mu} = \left(-1\right)^{p(n-p)+1} \tensor{\vartheta}{_\mu}$; see \cite{Hehl:2003}, Eq.\ (C.2.90). Since the coframe $\tensor{\vartheta}{_\mu}$ is a 1-form, we have $p=1$, and we are in four dimensions, that is, $n=4$. Therefore, $\star \star \tensor{\vartheta}{_\mu} = \tensor{\vartheta}{_\mu}$.

Applying the interior product to this 1-form then yields the metric tensor, $\tensor{e}{_\mu} \righthalfcup \tensor{\vartheta}{_\nu} = \tensor{g}{_\mu_\nu}$, because frame $\tensor{e}{_\mu}$ and coframe $\tensor{\vartheta}{_\nu}$ are dual to each other. This yields
\begin{align}
\tensor{e}{_\mu} \righthalfcup \star \tensor{\tilde{\Sigma}}{_\nu_\rho_\sigma} = \tensor{\text{Weyl}}{_\nu_\tau_\alpha_\rho}\tensor{\text{Weyl}}{_\mu^\tau^\alpha_\sigma} + \tensor{\text{Weyl}}{_\nu_\tau_\alpha_\sigma}\tensor{\text{Weyl}}{_\mu^\tau^\alpha_\rho} - \frac{1}{2} \tensor{g}{_\mu_\nu} \tensor{\text{Weyl}}{_\omega_\tau_\alpha_\rho}\tensor{\text{Weyl}}{^\omega^\tau^\alpha_\sigma} . \label{eq:bel_robinson_3_form_components}
\end{align}
The definition of the Bel tensor, see Eq.\ \eqref{eq:bel_tensor_definition_duals}, is equivalent to
\begin{align}
\begin{split}
\tensor{B}{_\mu_\nu_\rho_\sigma} =~ & \tensor{\text{Riem}}{_\mu_\alpha_\beta_\rho}\tensor{\text{Riem}}{_\nu^\alpha^\beta_\sigma} + \tensor{\text{Riem}}{_\mu_\alpha_\beta_\sigma}\tensor{\text{Riem}}{_\nu^\alpha^\beta_\rho} + \frac{1}{8}\tensor{g}{_\mu_\nu}\tensor{g}{_\rho_\sigma}\tensor{\text{Riem}}{_\alpha_\beta_\gamma_\delta}\tensor{\text{Riem}}{^\alpha^\beta^\gamma^\delta} \\
								 & - \frac{1}{2}\tensor{g}{_\mu_\nu} \tensor{\text{Riem}}{_\alpha_\beta_\gamma_\rho} \tensor{\text{Riem}}{^\alpha^\beta^\gamma_\sigma} - \frac{1}{2}\tensor{g}{_\rho_\sigma} \tensor{\text{Riem}}{_\alpha_\beta_\gamma_\mu} \tensor{\text{Riem}}{^\alpha^\beta^\gamma_\nu} . \label{eq:bel_tensor_components}
\end{split}
\end{align}
It coincides with the definition of the Bel--Robinson tensor \eqref{eq:bel_robinson_definition} when substituting the Weyl tensor for the Riemann tensor (in fact, it gives twice the Bel--Robinson tensor, rooted in these properties of the Weyl tensor: its left and right dual coincide, and its double dual is again the Weyl tensor, up to a sign).

Inserting the vacuum relation \eqref{eq:weyl_tensor_vacuum_relation} into the last summand of Eq.\ \eqref{eq:bel_tensor_components} then yields
\begin{align}
\tensor{\tilde{B}}{_\mu_\nu_\rho_\sigma} = \tensor{\text{Weyl}}{_\mu_\alpha_\beta_\rho}\tensor{\text{Weyl}}{_\nu^\alpha^\beta_\sigma} + \tensor{\text{Weyl}}{_\mu_\alpha_\beta_\sigma}\tensor{\text{Weyl}}{_\nu^\alpha^\beta_\rho} - \frac{1}{2}\tensor{g}{_\mu_\nu} \tensor{\text{Weyl}}{_\alpha_\beta_\gamma_\rho} \tensor{\text{Weyl}}{^\alpha^\beta^\gamma_\sigma} . \label{eq:bel_robinson_final}
\end{align}
The relation \eqref{eq:bel_robinson_final} is also found in the literature, see Bel \cite{Bel:1962}, Eq.\ (15), Robinson \cite{Robinson:1997}, Eq.\ (3.1), Garecki \cite{Garecki:2000dj}, Eq.\ (1), Douglas \cite{Douglas:2003}, Eq.\ (19), and So \cite{So:2010yq}, Eq.\ (1). It coincides with Eq.\ \eqref{eq:bel_robinson_3_form_components} and hence the proof is concluded: The Bel--Robinson tensor can indeed be expressed as
\begin{align}
\tensor{\tilde{B}}{_\mu_\nu_\rho_\sigma} = \tensor{e}{_\mu} \righthalfcup \star \tensor{\tilde{\Sigma}}{_\nu_\rho_\sigma} .
\end{align}

The Bel--Robinson 3-form is tracefree:
\begin{align}
\begin{split}
\tensor{\vartheta}{^\alpha} \wedge \tensor{\widetilde{\Sigma}}{_\alpha_\rho_\sigma} &= \left[ 2 \tensor{\text{Weyl}}{_\omega_\tau_\rho_\beta} \tensor{\text{Weyl}}{^\omega^\tau^\beta_\sigma} - \left( \tensor{\text{Weyl}}{_\alpha_\tau_\rho_\beta}\tensor{\text{Weyl}}{^\omega^\tau^\beta_\sigma} + \tensor{\text{Weyl}}{_\alpha_\tau_\sigma_\beta}\tensor{\text{Weyl}}{^\omega^\tau^\beta_\rho} \right) \tensor*{\delta}{^\alpha_\omega} \right] \eta \\
																					&= 2 \tensor{\text{Weyl}}{_\omega_\tau_\alpha_[_\sigma} \tensor{\text{Weyl}}{^\omega^\tau^\alpha_\rho_]} \eta \\
																					&= 0
\end{split}
\end{align}
In the first line, we used the identity $\tensor{\vartheta}{^\mu} \wedge \tensor{\eta}{_\nu} = \tensor*{\delta}{^\mu_\nu} \eta$, and in the last line we used the vacuum relation \eqref{eq:weyl_tensor_vacuum_relation}. We now show the equivalence of our result to the 3-form introduced by G\'omez-Lobo \cite{Gomez-Lobo:2014xsa}:
\begin{align}
\tensor{\mathcal{T}}{_\nu_\rho_\sigma} := \left[ \frac{1}{2}\tensor{g}{_\rho_\sigma} \star \left( \tensor{\text{Weyl}}{^\alpha^\beta} \wedge \tensor{\vartheta}{_\nu} \right) + 2 \tensor*{\delta}{^\beta_(_\rho} \star \left( \tensor{\text{Weyl}}{_\sigma_)^\alpha} \wedge \tensor{\vartheta}{_\nu} \right) \right] \wedge \tensor{\text{Weyl}}{_\alpha_\beta}
\end{align}
Expanding the above in components leads to
\begin{align}
\tensor{\mathcal{T}}{_\nu_\rho_\sigma} = \left( \frac{1}{8} \tensor{g}{_\rho_\sigma} \tensor{\text{Weyl}}{^\gamma^\delta^\alpha^\beta} \tensor{\text{Weyl}}{_\lambda_\epsilon_\alpha_\beta} \, \tensor{\eta}{_\gamma_\delta_\nu} + \frac{1}{2} \tensor*{\delta}{^\beta_(_\rho} \tensor{\text{Weyl}}{^\gamma^\delta_\sigma_)^\alpha} \tensor{\text{Weyl}}{_\epsilon_\lambda_\alpha_\beta} \, \tensor{\eta}{_\gamma_\delta_\nu} \right) \wedge \tensor{\vartheta}{^\epsilon} \wedge \tensor{\vartheta}{^\lambda} .
\end{align}
We now employ the identity $\tensor{\vartheta}{^\epsilon} \wedge \tensor{\vartheta}{^\lambda} \wedge \tensor{\eta}{_\gamma_\delta_\nu} = 3! \, \tensor*{\delta}{^\epsilon_[_\gamma} \tensor*{\delta}{^\lambda_\delta} \tensor{\eta}{_\nu_]}$ and obtain
\begin{align}
\begin{split}
\tensor{\mathcal{T}}{_\nu_\rho_\sigma} =\quad &\left( \frac{1}{4}\tensor{g}{_\rho_\sigma} \tensor{\text{Weyl}}{_\alpha_\beta_\gamma_\delta}\tensor{\text{Weyl}}{^\alpha^\beta^\gamma^\delta} - \tensor{\text{Weyl}}{_\gamma_\delta_\alpha_\rho}\tensor{\text{Weyl}}{^\gamma^\delta^\alpha_\sigma} \right) \tensor{\eta}{_\nu}\\
+ &\left( -\frac{1}{2} \tensor{g}{_\rho_\sigma}\tensor{\text{Weyl}}{^\alpha^\beta^\gamma^\delta}\tensor{\text{Weyl}}{_\alpha_\beta_\gamma_\nu} + \tensor{\text{Weyl}}{_\sigma^\alpha^\gamma^\delta}\tensor{\text{Weyl}}{_\nu_\gamma_\alpha_\rho} + \tensor{\text{Weyl}}{_\rho^\alpha^\gamma^\delta}\tensor{\text{Weyl}}{_\nu_\gamma_\alpha_\sigma} \right) \tensor{\eta}{_\delta} .
\end{split}
\end{align}
By means of the vacuum relation \eqref{eq:weyl_tensor_vacuum_relation} the first parenthesis vanishes and the second parenthesis turns out to be the Bel--Robinson 3-form as expressed in Eq.\ \eqref{eq:bel_robinson_3_form_intermediate}:
\begin{align}
\tensor{\mathcal{T}}{_\nu_\rho_\sigma} =\frac{1}{2} \tensor{\text{Weyl}}{_\beta_\gamma_\rho_\alpha} \tensor{\text{Weyl}}{^\beta^\gamma^\alpha_\sigma} \tensor{\eta}{_\nu} - \left( \tensor{\text{Weyl}}{_\nu_\gamma_\rho_\alpha}\tensor{\text{Weyl}}{^\delta^\gamma^\alpha_\sigma} + \tensor{\text{Weyl}}{_\nu_\gamma_\sigma_\alpha}\tensor{\text{Weyl}}{^\delta^\gamma^\alpha_\rho} \right) \tensor{\eta}{_\delta} .
\end{align}

\section{Exterior calculus}
\label{appendix:exterior_calculus}
The following is a brief outline of our notation in exterior calculus.

For a Riemannian spacetime, the \emph{anholonomic coframe} is given by $\tensor{\vartheta}{^\mu} = \tensor{e}{_a^\mu}\tensor{dx}{^a}$ in terms of the \emph{holonomic coordinate cobasis} $\tensor{dx}{^i}$. Similarly, the \emph{anholonomic frame} is $\tensor{e}{_\mu} = \tensor{e}{^a_\mu} \tensor{\partial}{_a}$, where $\tensor{\partial}{_i}$ is the holonomic coordinate basis. The expansion coefficients $\tensor{e}{_a^\mu}$ are called the \emph{tetrad}. Frame and coframe are \emph{dual} to each other, that is, $\tensor{e}{_\nu} \righthalfcup \tensor{\vartheta}{^\mu} = \tensor*{\delta}{^\mu_\nu}$, where $\righthalfcup$ denotes the \emph{interior product}. We use Greek indices for anholonomic frame components and Latin indices for holonomic coordinate components.

The \emph{metric} g is introduced as the symmetric \tvect{0}{2} tensor field $g = \tensor{g}{_a_b} \tensor{dx}{^a} \otimes \tensor{dx}{^b} = \tensor{g}{_\alpha_\beta} \tensor{\vartheta}{^\alpha} \otimes \tensor{\vartheta}{^\beta}$. $\tensor{g}{_i_j}$ is used to raise and lower coordinate indices, and $\tensor{g}{_\mu_\nu}$ applies to anholonomic indices. We use the degree of freedom granted by the tetrad to set $\left( \tensor{g}{_\mu_\nu} \right) = \text{diag} \, \left( -1, 1, 1, 1\right)$, thereby enforcing a \emph{pseudo-orthonormal} coframe $\tensor{\vartheta}{^\mu}$.

After the appearance of the metric, the Hodge dual ``$\star$'' can be introduced, mapping $p$-forms to $(n-p)$-forms. We introduce the \emph{$\eta$-basis}:
\begin{alignat}{8}
\eta \quad &:=&& \quad \star 1 & && \text{4-form} \nonumber \\
\tensor{\eta}{_\mu} \quad &:=&& \quad \tensor{e}{_\mu} \righthalfcup \eta \quad &=& \quad \star \left( \tensor{\vartheta}{_\mu} \right) \qquad & \text{3-form} \nonumber \\
\tensor{\eta}{_\mu_\nu} \quad &:=&& \quad \tensor{e}{_\nu} \righthalfcup \tensor{\eta}{_\mu} \quad &=& \quad \star \left( \tensor{\vartheta}{_\mu} \wedge \tensor{\vartheta}{_\nu} \right) \qquad & \text{2-form} \label{eq:eta-basis} \\
\tensor{\eta}{_\mu_\nu_\rho} \quad &:=&& \quad \tensor{e}{_\rho} \righthalfcup \tensor{\eta}{_\mu_\nu} \quad &=& \quad \star \left( \tensor{\vartheta}{_\mu} \wedge \tensor{\vartheta}{_\nu} \wedge \tensor{\vartheta}{_\rho} \right) \qquad & \text{1-form} \nonumber \\
\tensor{\eta}{_\mu_\nu_\rho_\sigma} \quad &:=&& \quad \tensor{e}{_\sigma} \righthalfcup \tensor{\eta}{_\mu_\nu_\rho} \quad &=& \quad \star \left( \tensor{\vartheta}{_\mu} \wedge \tensor{\vartheta}{_\nu} \wedge \tensor{\vartheta}{_\rho} \wedge \tensor{\vartheta}{_\sigma} \right) \qquad & \text{0-form} \nonumber
\end{alignat}
$\wedge$ denotes the \emph{exterior product} of forms and $\tensor{\vartheta}{_\mu} = \tensor{g}{_\mu_\alpha}\tensor{\vartheta}{^\alpha}$. The Hodge dual acts on a $p$-form $\omega$ as follows, mapping it to an $(n-p)$-form:
\begin{align}
\star \, \omega = \star \, \left( \frac{1}{p!} \tensor{\omega}{_{\alpha_1}_\dots_{\alpha_p}} \tensor{\vartheta}{^{\alpha_1}} \wedge \dots \wedge \tensor{\vartheta}{^{\alpha_p}} \right) := \frac{1}{p!(n-p)!}\tensor{\omega}{_{\alpha_1}_\dots_{\alpha_p}} \tensor{\eta}{^{\alpha_1}^\dots^{\alpha_p}_{\beta_{p+1}\dots\beta_n}} \, \tensor{\vartheta}{^{\beta_{p+1}}} \wedge \dots \wedge \tensor{\vartheta}{^{\beta_n}}
\end{align}
$\tensor{\eta}{_\mu_\nu_\rho_\sigma}$ is the totally antisymmetric unit tensor. It can be used to define a \emph{tensor dual} acting on $p \in [0, n]$ antisymmetric indices. For a tensor $\tensor{T}{_\mu_\nu_\alpha_\beta}$ satisfying $\tensor{T}{_(_\mu_\nu_)_\alpha_\beta} = \tensor{T}{_\mu_\nu_(_\alpha_\beta_)} = 0$, we define the \emph{left}, \emph{right}, and \emph{double tensor dual} according to
\begin{align}
\begin{split}
\tensor{\text{*}T}{_\kappa_\lambda_\alpha_\beta} & := \frac{1}{2} \tensor{\eta}{_\kappa_\lambda^\mu^\nu} \tensor{T}{_\mu_\nu_\alpha_\beta} , \\
\tensor{T\text{*}}{_\mu_\nu_\rho_\sigma} & := \frac{1}{2} \tensor{T}{_\mu_\nu_\alpha_\beta} \tensor{\eta}{^\alpha^\beta_\rho_\sigma} , \\
\tensor{\text{*}T\text{*}}{_\kappa_\lambda_\rho_\sigma} & := \frac{1}{4} \tensor{\eta}{_\kappa_\lambda^\mu^\nu} \tensor{T}{_\mu_\nu_\alpha_\beta} \tensor{\eta}{^\alpha^\beta_\rho_\sigma} .
\end{split}
\end{align}
In a pseudo-orthonormal coframe, the metric compatible, torsion free Levi--Civita \emph{connection} 1-form $\tensor{\Gamma}{_\mu_\nu} = \tensor{\Gamma}{_\alpha_\mu_\nu}\,\tensor{\vartheta}{^\alpha}$ is antisymmetric $\tensor{\Gamma}{_\mu_\nu} = -\tensor{\Gamma}{_\nu_\mu}$. In Riemannian geometry, it is completely determined by the coframe:
\begin{align}
\tensor{\Gamma}{_\mu_\nu} = \frac{1}{2} \left( \tensor{e}{_\mu} \righthalfcup \tensor{e}{_\nu} \righthalfcup \tensor{\Omega}{_\alpha} \right) \tensor{\vartheta}{^\alpha} -\tensor{e}{_[_\mu} \righthalfcup \tensor{\Omega}{_\nu_]} + \tensor{e}{_{\{}_\nu} \tensor{g}{_\alpha_\mu_{\}}} \tensor{\vartheta}{^\alpha} \label{eq:connection}
\end{align}
$\{\alpha\beta\gamma\} := \frac{1}{2}\left( \alpha\beta\gamma + \beta\gamma\alpha - \gamma\alpha\beta\right)$ is the so-called \emph{Schouten bracket} and $\tensor{\Omega}{^\mu} := d \tensor{\vartheta}{^\mu} = \frac{1}{2}\,\tensor{\Omega}{_\alpha_\beta^\mu}\,\tensor{\vartheta}{^\alpha}\wedge\tensor{\vartheta}{^\beta}$ is the \emph{object of anholonomity} 2-form and expresses the failure of the anholonomic frame to be integrable. Its components are given by
\begin{align}
\tensor{\Omega}{_\rho_\sigma^\mu} = 2 \tensor{e}{^a_\rho} \tensor{e}{^b_\sigma} \tensor{\partial}{_[_a} \tensor{e}{_b_]^\mu} \quad .
\end{align}
Therefore, the object of anholonomity vanishes identically for holonomic coordinate bases since there the tetrads are simply given by $\tensor{e}{_i^\mu} = \tensor*{\delta}{^\mu_i} = const$. Note that in such a basis, Eq.\ (\ref{eq:connection}) reduces to the well-known formula for the Levi-Civita connection (or \emph{Christoffel symbols}, as they are mostly referred to within tensor calculus):
\begin{align}
\tensor{\Gamma}{_i_j} = 0 - 0 + \tensor{\partial}{_{\{}_j} \tensor{g}{_k_i_{\}}} \tensor{dx}{^k} = \frac{1}{2} \left( \tensor{\partial}{_j} \tensor{g}{_k_i} + \tensor{\partial}{_k} \tensor{g}{_i_j} - \tensor{\partial}{_i} \tensor{g}{_j_k} \right) \, \tensor{dx}{^k} = \tensor{\Gamma}{_k_i_j} \tensor{dx}{^k}
\end{align}
We now have a \emph{covariant derivative} at our disposal, acting on a $p$-form $\tensor{T}{^\mu_\nu}$ as
\begin{align}
D\tensor{T}{^\mu_\nu} := d\tensor{T}{^\mu_\nu} + \tensor{\Gamma}{^\mu_\alpha} \wedge \tensor{T}{^\alpha_\nu} - \tensor{\Gamma}{^\alpha_\nu} \wedge \tensor{T}{^\mu_\alpha} \quad .
\end{align}
$d$ denotes the \emph{exterior derivative} of forms, and $\wedge$ is the \emph{exterior product}.

The antisymmetric \emph{curvature} 2-form is given by $\tensor{\text{Riem}}{^\mu_\nu} := d\tensor{\Gamma}{^\mu_\nu} + \tensor{\Gamma}{^\mu_\alpha} \wedge \tensor{\Gamma}{^\alpha_\nu} = \frac{1}{2} \tensor{\text{Riem}}{_\alpha_\beta^\mu_\nu} \tensor{\vartheta}{^\alpha} \wedge \tensor{\vartheta}{^\beta}$. Its contraction, the Ricci 1-form, reads $\tensor{\text{Ric}}{_\mu} := \tensor{e}{_\alpha} \righthalfcup \tensor{\text{Riem}}{^\alpha_\mu} = \tensor{\text{Ric}}{_\alpha_\mu}\tensor{\vartheta}{^\alpha}$. The Ricci scalar is then $\text{R} := \tensor{e}{_\alpha} \righthalfcup \tensor{\text{Ric}}{^\alpha}$. The \emph{Einstein equation} with \emph{cosmological constant} $\Lambda$ and energy-momentum 3-form $\tensor{\Sigma}{_\mu}$ then takes the form
\begin{align}
\tensor{G}{_\mu} + \Lambda\tensor{\eta}{_\mu} = 8 \pi \tensor{\Sigma}{_\mu} . \label{eq:einstein_equation}
\end{align}
$\tensor{G}{_\mu} := \frac{1}{2}\tensor{\eta}{_\mu_\alpha_\beta} \wedge \tensor{\text{Riem}}{^\alpha^\beta}$ is the \emph{Einstein} 3-form. It comes about somewhat disguised, but its dual turns out to be $\star \tensor{G}{_\mu} = \tensor{\text{Ric}}{_\mu} - \frac{1}{2}\text{R}\tensor{\vartheta}{_\mu}$, which can be translated directly into the anholonomic, 0-form valued components of the Einstein tensor known from tensor calculus: $\tensor{G}{_\mu_\nu} := \tensor{e}{_\mu} \righthalfcup \star \tensor{G}{_\nu}$. Using the dual of the Einstein 3-form, we may write the Einstein equations as
\begin{align}
\tensor{\text{Ric}}{_\mu} - \frac{1}{2} \text{R} \tensor{\vartheta}{_\mu} + \Lambda \tensor{\vartheta}{_\mu} = 8\pi \star \tensor{\Sigma}{_\mu} . \label{eq:einstein_equation_dual}
\end{align}

\section{Computer algebra code}
\label{appendix:computer_algebra}
The computer algebra system Reduce \cite{Hearn} was employed to carry out the calculations, supplemented by the package Excalc providing an efficient framework for exterior calculus. The source codes are listed below. For a review of the computer algebra system Reduce with Excalc applied to general relativity and beyond, see e.g.\ Socorro \textit{et al.} \cite{Socorro:1998hr}.

\subsection{Pleba\'nski--Demia\'nski coframe and vector potential}
\label{appendix:pb_check}
The listing below first defines the Pleba\'nski--Demia\'nski coframe and its accompanying vector potential. Then a variety of standard programs is included (see appendix \ref{appendix:code_invariants}) for decomposing the curvature, calculating invariants, and calculating various tensorial quantities. Thereby the validity of the solution can easily be checked. This is not only reasonable for itself, it also serves the purpose to exclude any inconsistencies in our own notation. In a third step, the coordinate transformations and rescalings introduced by Griffiths and Podolsk\'y \cite{Griffiths:2005qp} are checked for consistency.
\vspace{15pt}
\begin{singlespace}
\begin{lstlisting}[title=file: plebanski\_demianski\_v6.rei]
%%%%%%%%%%%%%%%%%%%%%%%%%%%%%%%%%%%%%%%%%%%%%%%%%
%
% REDUCE file checking the Plebanski-Demianski
% solution, and calculating second order
% curvature invariants within exterior calculus
%
% last edited by J. Boos, Jan 7, 2015
%
% file: plebanski_demianski_v6.rei
%
% conventions: 05_elm_inv_v6.pdf
%
%%%%%%%%%%%%%%%%%%%%%%%%%%%%%%%%%%%%%%%%%%%%%%%%%


%%%%%%%%%%%%%%%%%%%%%%%%%%%%%%%%%%%%%%%%%%%%%%%%%
%
% load the package excalc for exterior calculus
% and adjust the line length of the output
%

	load_package excalc $
	linelength(200) $


%%%%%%%%%%%%%%%%%%%%%%%%%%%%%%%%%%%%%%%%%%%%%%%%%
%
% define the frame
%
%  constants: c_m ......... mass-like parameter
%             c_n ......... NUT-like parameter
%             c_epsilon.... ?
%             c_gamma...... ?
%             c_e ......... electric charge
%             c_g ......... (hypothetical) magnetic charge
%             c_lambda..... cosmological constant
%

	% auxiliary functions
	clear {c_qq, c_pp, c_delta, c_hh, sqrt_c_qq, sqrt_c_pp} $
	pform {c_qq, c_pp, c_delta, c_hh, sqrt_c_qq, sqrt_c_pp} = 0 $
	
	% set the domains
	fdomain c_qq = c_qq(q), sqrt_c_qq = sqrt_c_qq(q),
			c_pp = c_pp(p), sqrt_c_pp = sqrt_c_pp(p), 
			c_delta = c_delta(p, q), c_hh = c_hh(p, q) $
	
	% Plebanski-Demianski coframe
	coframe o(0) = 1/c_hh*sqrt_c_qq/sqrt(c_delta)*(d tau - p**2*d sigma ) ,
	        o(1) = 1/c_hh*sqrt(c_delta)/sqrt_c_qq*d q ,	        
	        o(2) = cv_sign/c_hh*sqrt(c_delta)/sqrt_c_pp*d p ,
	        o(3) = cv_sign/c_hh*sqrt_c_pp/sqrt(c_delta)*(d tau + q**2*d sigma )
	with metric g = -o(0)*o(0) + o(1)*o(1) + o(2)*o(2) + o(3)*o(3) $

	% conventional sign, see 05_elm_inv_v5.pdf for details
	cv_sign := -1 $

	% specify the functions explicitly, except for the quartics c_qq, c_pp
	c_delta := p**2 + q**2 $
	c_hh := 1 - p*q $
	
	% specify the chain rule for the square root expressions
	@(sqrt_c_qq, q) := @(c_qq, q)/(2*sqrt_c_qq) $
	@(sqrt_c_pp, p) := @(c_pp, p)/(2*sqrt_c_pp) $
	sqrt_c_qq**2 := c_qq $
	sqrt_c_pp**2 := c_pp $

	% denote the frame by e
	frame e $

	% show the frame as a test
	displayframe $
	
	% vector potential
	clear a1 $
	pform a1 = 1 $
	a1 := c_hh/sqrt(c_delta)*( c_e*q/sqrt_c_qq*o(0) + c_g*p/sqrt_c_pp*o(3) ) $


%%%%%%%%%%%%%%%%%%%%%%%%%%%%%%%%%%%%%%%%%%%%%%%%%
%
% calculate quantities
%

	in "einstein_maxwell_v1.rei" $
	in "curvature_v1.rei" $
	in "invariants_v1.rei" $
	in "weyl_def_v2.rei" $
	in "newman_penrose_v1.rei" $
	in "matte_v1.rei" $
	in "bel_v1.rei" $
	in "bel_robinson_v2.rei" $
	in "kummer_v1.rei" $

		
%%%%%%%%%%%%%%%%%%%%%%%%%%%%%%%%%%%%%%%%%%%%%%%%%
%
% final substitution of the quartics,
% various checks
%

	% quartic functions
	c_pp := c_k + 2*c_n*p - c_epsilon*p**2 + 2*c_m*p**3 - (c_k + c_e**2 + c_g**2 - c_lambda/3)*p**4 $
	c_qq := c_k + c_e**2 + c_g**2 - 2*c_m*q + c_epsilon*q**2 - 2*c_n*q**3 - (c_k - c_lambda/3)*q**4 $
	sqrt_c_pp := sqrt(c_pp) $
	sqrt_c_qq := sqrt(c_qq) $

	% solution of Einstein-Maxwell equations?
	% (this expression should vanish identically)
	write emtest3(a) := emtest3(a) $
	
	% solution of Maxwell equations?
	write maxhom3 := maxhom3 $
	write maxinhom3 := maxinhom3 $
	
	% define shorthands as they appear in the curvature
	ee := -1/2*weyl0(-0,-1,-0,-1) $
	bb := 1/2*weyl0(-0,-1,-2,-3) $
	qq := -2*# (o(0) ^ sigma3(-0)) $
	
	% find greatest common divisor
	on gcd $
	
	% Kretschmann invariants, also check decomposition
	write kretschmannw0 / (bb**2 - ee**2) $ 
	write kretschmannr0 / qq**2 $
	write kretschmanns0 / c_lambda**2 $
	write kretschmann0 - kretschmannw0 - kretschmannr0 - kretschmanns0 $
	
	% Pontryagin pseudo-invariants, also check decomposition
	write pontryaginw0 / ee / bb $ 
	write pontryaginr0 $
	write pontryagins0 $
	write pontryagin0 - pontryaginw0 - pontryaginr0 - pontryagins0 $
	
	% greatest common divisor not always needed in the following
	off gcd $
	
	% check the expressions for ee and bb
	ee_test := ((p*q-1)/(p**2 + q**2))**3*((3*p**2 - q**2)*c_m*q + (p**2 - 3*q**2)*c_n*p - (c_e**2 + c_g**2)*(1 + p*q)*(p**2 - q**2)) $
	bb_test := ((p*q-1)/(p**2 + q**2))**3*((p**2 - 3*q**2)*c_m*p - (3*p**2 - q**2)*c_n*q + 2*(c_e**2 + c_g**2)*(1 + p*q)*p*q) $
	write ee - ee_test $
	write bb - bb_test $

	end $

%%%%%%%%%%%%%%%%%%%%%%%%%%%%%%%%%%%%%%%%%%%%%%%%%
%
% introduce coordinates of Griffiths & Podolsky
%

	% transform coordinates
	q := sqrt(c_alpha/c_w)*r $
	p := sqrt(c_alpha/c_w)*(c_l + c_a*cos(theta)) $	
	c_m := (c_alpha/c_w)**(3/2)*c_gp_m $
	c_n := (c_alpha/c_w)**(3/2)*c_gp_n $
	c_e := c_alpha/c_w*c_gp_e $
	c_g := c_alpha/c_w*c_gp_g $
	c_epsilon := c_alpha/c_w*c_gp_epsilon $
	c_k := c_alpha**2*c_gp_k $
	c_lambda := c_gp_lambda $
	
	% constrain old, free parameters in terms of the new parameters
	c_gp_epsilon := c_w**2*c_gp_k2 + 4*c_alpha/c_w*c_l*c_gp_m - (c_a**2 + 3*c_l**2)*(c_alpha**2/c_w**2*(c_w**2*c_gp_k + c_gp_e**2 + c_gp_g**2) - c_gp_lambda/3) $
	c_gp_n := c_w**2*c_gp_k2*c_l - c_alpha*(c_a**2 - c_l**2)*c_gp_m/c_w + (c_a**2 - c_l**2)*c_l*(c_alpha**2/c_w**2*(c_w**2*c_gp_k + c_gp_e**2 + c_gp_g**2) - c_gp_lambda/3) $	
	c_gp_k	:= (c_a**2 - c_l**2)*(1 + 2*c_alpha*c_l*c_gp_m/c_w - 3*c_alpha**2*c_l**2/c_w**2*(c_gp_e**2 + c_gp_g**2) + c_l**2*c_gp_lambda)/(c_w**2 + 3*c_alpha**2*c_l**2*(c_a**2 - c_l**2)) $
	c_gp_k2	:= (1 + 2*c_alpha*c_l*c_gp_m/c_w - 3*c_alpha**2*c_l**2/c_w**2*(c_gp_e**2 + c_gp_g**2) + c_l**2*c_gp_lambda)/(c_w**2 + 3*c_alpha**2*c_l**2*(c_a**2 - c_l**2)) $
	
	% auxiliary functions
	aux_a3 := 2*c_alpha*c_a*c_gp_m/c_w - 4*c_alpha**2*c_a*c_l/c_w**2*(c_w**2*c_gp_k + c_gp_e**2 + c_gp_g**2) + 4/3*c_gp_lambda*c_a*c_l $
	aux_a4 := -c_alpha**2*c_a**2/c_w**2*(c_w**2*c_gp_k + c_gp_e**2 + c_gp_g**2) + c_gp_lambda/3*c_a**2 $
	
	% check the transformations of the abbreviations
	% (all of these expressions yield zero)
	
	c_gp_delta := c_w**2*c_gp_k + c_gp_e**2 + c_gp_g**2 - 2*c_gp_m*r + c_gp_epsilon*r**2 - 2*c_alpha/c_w*c_gp_n*r**3 - (c_alpha**2*c_gp_k - c_gp_lambda/3)*r**4 $
	write c_gp_delta - c_w**2/c_alpha**2*c_qq $ 
	
	c_gp_chi := 1 - aux_a3*cos(theta) - aux_a4*cos(theta)**2 $
	write c_gp_chi - c_pp*c_w**2/c_alpha**2/c_a**2/(1-cos(theta)**2) $
	
	c_gp_rho := sqrt(r**2 + (c_l + c_a*cos(theta))**2) $
	write c_gp_rho**2 - c_w/c_alpha*(p**2 + q**2) $
	
	c_omega := 1 - c_alpha/c_w*r*(c_l + c_a*cos(theta)) $
	write c_omega - (1 - p*q) $

end $
\end{lstlisting}
\end{singlespace}

\subsection{Griffiths--Podolsk\'y coframe and vector potential}
\label{appendix:gp_check}
For the sake of completeness, we also checked the expressions for the components of the Riemann and Weyl tensors as well as the invariants directly in the Griffiths--Podolsk\'y coframe. These calculations are quite time-consuming, even after several optimizations. We assume that this is due to the extensive redefinitions of the original constants and the non-polynomial structure of the coframe in the GP coordinates.

The Bel, Bel--Robinson, and Kummer tensors are not evaluated again, since their structure follows algebraically from the confirmed structure of the Riemann and Weyl tensor.
\vspace{15pt}
\begin{singlespace}
\begin{lstlisting}[title=file: griffiths\_podolsky\_v5.rei]
%%%%%%%%%%%%%%%%%%%%%%%%%%%%%%%%%%%%%%%%%%%%%%%%%
%
% REDUCE file checking the coordinates of
% Griffiths & Podolsky for the Plebanski-
% Demianski solution
%
% last edited by J. Boos, Jan 7, 2014
%
% file: griffiths_podolsky_v5.rei
%
% conventions: 05_elm_inv_v6.pdf
%
%%%%%%%%%%%%%%%%%%%%%%%%%%%%%%%%%%%%%%%%%%%%%%%%%


%%%%%%%%%%%%%%%%%%%%%%%%%%%%%%%%%%%%%%%%%%%%%%%%%
%
% load the package excalc for exterior calculus
% and adjust the line length of the output
%

	load_package excalc $
	linelength(200) $


%%%%%%%%%%%%%%%%%%%%%%%%%%%%%%%%%%%%%%%%%%%%%%%%%
%
% define the frame and vector potential
%
%  constants: c_m ......... mass
%             c_a ......... angular momentum per mass
%             c_alpha...... acceleration parameter
%             c_lambda .... cosmological constant
%             c_l ......... Taub-NUT parameter
%             c_e ......... electric charge
%             c_g ......... (hypothetical) magnetic charge
%
%             c_w.......... scaling degree of freedom,
%                           here set to unity (see below)
%

	% auxiliary functions
	clear c_delta, c_rho, c_chi, c_omega $
	pform {c_delta, c_rho, c_chi, c_omega} = 0 $
	
	% auxiliary functions for some square root expressions
	clear c_sqrt_delta, c_sqrt_chi $
	pform {c_sqrt_delta, c_sqrt_chi} = 0 $
	
	% set their domain
	fdomain c_omega = c_omega(r, theta),
			c_delta = c_delta(r),
			c_rho = c_rho(r, theta),
			c_chi = c_chi(theta),
			c_sqrt_delta = c_sqrt_delta(r),
			c_sqrt_chi = c_sqrt_chi(theta) $
	
	% specify chain rule implementation for the square roots
	@(c_sqrt_delta, r) := @(c_delta, r)/(2*c_sqrt_delta) $
	@(c_sqrt_chi, theta) := @(c_chi, theta)/(2*c_sqrt_chi) $
	c_sqrt_delta**2 := c_delta $
	c_sqrt_chi**2 := c_chi $
	@(c_rho, r) := r/c_rho $
	@(c_rho, theta) := -c_a*sin(theta)*(c_l + c_a*cos(theta))/c_rho $

	% coframe ( employed half angle formula for sin**2(theta/2) )
	coframe o(0) = 1/c_omega*c_sqrt_delta/c_rho*(d t - (c_a*sin(theta)**2 - 2*c_l*cos(theta) + 2*c_l)* d phi ) ,
	        o(1) = 1/c_omega*c_rho/c_sqrt_delta* d r ,
	        o(2) = -cv_sign/c_omega*c_rho/c_sqrt_chi*d theta ,
	        o(3) = cv_sign/c_omega*sin(theta)*c_sqrt_chi/c_rho*(c_a*d t - (r**2 + (c_a+c_l)**2)*d phi )
	with metric g = -o(0)*o(0) + o(1)*o(1) + o(2)*o(2) + o(3)*o(3)  $
	
	% conventional sign,
	% see 05_elm_inv_v3.pdf for details
	cv_sign := -1 $
	% cv_sign**2 := 1 $

	% set scaling dof to 1
	% c_w := 1 $
	
	% denote the frame by e
	frame e $

	% show the frame as a test
	displayframe $
	
	% vector potential
	clear a1 $
	pform a1 = 1 $
	a1 := c_e*c_omega/c_rho*r/c_sqrt_delta*o(0) + c_g*c_omega/c_rho*(c_l/c_a + cos(theta))/c_sqrt_chi/sin(theta)*o(3) $

	% express all occuring sines in terms of cosines
	for all x let sin(x)**2 = 1 - cos(x)**2 $


%%%%%%%%%%%%%%%%%%%%%%%%%%%%%%%%%%%%%%%%%%%%%%%%%
%
% calculate quantities
%

	in "einstein_maxwell_v1.rei" $
	in "curvature_v1.rei" $
	in "invariants_v1.rei" $


%%%%%%%%%%%%%%%%%%%%%%%%%%%%%%%%%%%%%%%%%%%%%%%%%%
%
% introduce the constants and auxiliary
% functions at the end, this reduces computing time
%

	% define a procedure to update all important entities
	% (run this after each substitution, otherwise it takes much longer)
	procedure update $
		begin $
			emtest3(a) := emtest3(a) $
			sigma3(a) := sigma3(a) $
			kretschmann0 := kretschmann0 $
			kretschmannw0 := kretschmannw0 $
			kretschmannr0 := kretschmannr0 $
			kretschmanns0 := kretschmanns0 $
			pontryagin0 := pontryagin0 $
			pontryaginw0 := pontryaginw0 $
			pontryaginr0 := pontryaginr0 $
			pontryagins0 := pontryagins0 $		
		end $
	
	% auxiliary functions
	
	c_delta := c_w**2*aux_k + c_e**2 + c_g**2 - 2*c_m*r + aux_epsilon*r**2 - 2*c_alpha/c_w*aux_n*r**3 - (c_alpha**2*aux_k - c_lambda/3)*r**4 $ update () $
	c_chi := 1 - aux_a3*cos(theta) - aux_a4*cos(theta)**2 $	update () $
	c_rho := sqrt(r**2 + (c_l + c_a*cos(theta))**2) $ update () $
	c_omega := 1 - c_alpha/c_w*r*(c_l + c_a*cos(theta)) $ update () $
	c_sqrt_delta := sqrt(c_delta) $ update () $
	c_sqrt_chi := sqrt(c_chi) $ update () $
	
	% constants
	aux_a3 := 2*c_alpha*c_a*c_m/c_w - 4*c_alpha**2*c_a*c_l/c_w**2*(c_w**2*aux_k + c_e**2 + c_g**2) + 4/3*c_lambda*c_a*c_l $ update () $
	aux_a4 := -c_alpha**2*c_a**2/c_w**2*(c_w**2*aux_k + c_e**2 + c_g**2) + c_lambda/3*c_a**2 $ update () $
	aux_epsilon := c_w**2*aux_k2 + 4*c_alpha/c_w*c_l*c_m - (c_a**2 + 3*c_l**2)*(c_alpha**2/c_w**2*(c_w**2*aux_k + c_e**2 + c_g**2) - c_lambda/3) $ update () $
	aux_n := c_w**2*aux_k2*c_l - c_alpha*(c_a**2 - c_l**2)*c_m/c_w + (c_a**2 - c_l**2)*c_l*(c_alpha**2/c_w**2*(c_w**2*aux_k2*(c_a**2 - c_l**2) + c_e**2 + c_g**2) - c_lambda/3) $ update () $
	aux_k	:= (1 + 2*c_alpha*c_l*c_m/c_w - 3*c_alpha**2*c_l**2/c_w**2*(c_e**2 + c_g**2) + c_l**2*c_lambda)/(c_w**2/(c_a**2 - c_l**2) + 3*c_alpha**2*c_l**2) $
	aux_k2	:= (1 + 2*c_alpha*c_l*c_m/c_w - 3*c_alpha**2*c_l**2/c_w**2*(c_e**2 + c_g**2) + c_l**2*c_lambda)/(c_w**2 + 3*c_alpha**2*c_l**2*(c_a**2 - c_l**2)) $ update () $

	
%%%%%%%%%%%%%%%%%%%%%%%%%%%%%%%%%%%%%%%%%%%%%%%%%
%
% various checks
%
	
	% solution of Einstein-Maxwell equations?
	% (this expression should vanish identically)
	write emtest3(a) := emtest3(a) $
	
	% solution of Maxwell equations?
	write maxhom3 := maxhom3 $
	write maxinhom3 := maxinhom3 $
	
	% define shorthands as they appear in the curvature
	ee := -1/2*weyl0(-0,-1,-0,-1) $
	bb := 1/2*weyl0(-0,-1,-2,-3) $
	qq := -2*# (o(0) ^ sigma3(-0)) $
	
	% find greatest common divisor
	on gcd $
	
	% Kretschmann invariants, also check decomposition
	write kretschmannw0 / (bb**2 - ee**2) $ 
	write kretschmannr0 / qq**2 $
	write kretschmanns0 / c_lambda**2 $
	write kretschmann0 - kretschmannw0 - kretschmannr0 - kretschmanns0 $
	
	% Pontryagin pseudo-invariants, also check decomposition
	write pontryaginw0 / ee / bb $ 
	write pontryaginr0 $
	write pontryagins0 $
	write pontryagin0 - pontryaginw0 - pontryaginr0 - pontryagins0 $
	
	% greatest common divisor not always needed in the following
	off gcd $
	
	% check the expressions for ee and bb
	ee_test := c_omega**3/c_rho**6*((q**2 - 3*p**2)*c_m_hat*q + (3*q**2 - p**2)*c_n_hat*p - (c_e_hat**2 + c_g_hat**2)*(q**2 - p**2)*(1 + c_alpha*p*q)) $
	bb_test := c_omega**3/c_rho**6*((3*q**2 - p**2)*c_m_hat*p - (q**2 - 3*p**2)*c_n_hat*q - 2*(c_e_hat**2 + c_g_hat**2)*(1 + c_alpha*p*q)*p*q) $
	c_m_hat := c_m $
	c_n_hat := aux_n $
	c_e_hat := c_e $
	c_g_hat := c_g $
	p := c_l + c_a*cos(theta) $
	q := r $
	write ee - ee_test $
	write bb - bb_test $

end $
\end{lstlisting}
\end{singlespace}

\subsection{Universal programs for curvature decomposition, invariants, and other geometric objects}
\label{appendix:code_invariants}
The source codes below are a collection of standard code snippets that can be included in a Reduce program one after the other, once a coframe has been defined. For a suitable application, see the code above in appendix \ref{appendix:pb_check}.

\subsubsection{Check of Einstein--Maxwell equations}
\begin{singlespace}
\begin{lstlisting}[title=file: einstein\_maxwell\_v1.rei]
%%%%%%%%%%%%%%%%%%%%%%%%%%%%%%%%%%%%%%%%%%%%%%%%%
%
% REDUCE file
%
% purpose: checks if Einstein-Maxwell equations
% are fulfilled for coframe and vector potential
%
% last edited by J. Boos, Dec 3, 2014
%
% file: einstein_maxwell_v1.rei
%
% conventions: 05_elm_inv_v5.pdf
%
%%%%%%%%%%%%%%%%%%%%%%%%%%%%%%%%%%%%%%%%%%%%%%%%%

%%%%%%%%%%%%%%%%%%%%%%%%%%%%%%%%%%%%%%%%%%%%%%%%%%
%
% Levi-Civita connection for Riemannian geometry
%

	write "    connection..." $
	clear conx1 $
	pform conx1(a, b) = 1 $
	riemannconx conx1 $
	write "       done." $
	
%%%%%%%%%%%%%%%%%%%%%%%%%%%%%%%%%%%%%%%%%%%%%%%%%%
%
% eta basis
%

	write "    eta basis..." $
	clear eta0, eta1, eta2, eta3, eta4 $
	pform eta0(a, b, c, d) = 0 ,
	      eta1(a, b, c)    = 1 ,
	      eta2(a, b)       = 2 ,
	      eta3(a)          = 3 ,
	      eta4             = 4 $
	
	eta4             := # 1 $
	eta3(a)          := e(a) _| eta4 $
	eta2(a, b)       := e(b) _| eta3(a) $
	eta1(a, b, c)    := e(c) _| eta2(a, b) $
	eta0(a, b, c, d) := e(d) _| eta1(a, b, c) $
	write "       done." $

%%%%%%%%%%%%%%%%%%%%%%%%%%%%%%%%%%%%%%%%%%%%%%%%%%
%
% Riemann curvature 2-form
%

	write "    Riemann curvature..." $

	% curvature is an antisymmetric 2-form
	clear riem2 $
	pform riem2(a, b) = 2 $
	antisymmetric riem2 $
	riem2(a, -b) := d conx1(a, -b) + conx1(a, -c) ^ conx1(c, -b) $
	
%%%%%%%%%%%%%%%%%%%%%%%%%%%%%%%%%%%%%%%%%%%%%%%%%%
%
% Einstein 3-form
%

	write "    Einstein 3-form..." $

	% Einstein 3-form
	clear einstein3 $
	pform einstein3(a) = 3 $
	einstein3(-a) := 1/2 * eta1(-a, -b, -c) ^ riem2(b, c) $
	
	% Einstein tensor components
	clear einstein0 $
	pform einstein0(a, b) = 0 $
	einstein0(b, a) := e(b) _| ( # einstein3(a) ) $

	write "        done." $

%%%%%%%%%%%%%%%%%%%%%%%%%%%%%%%%%%%%%%%%%%%%%%%%%%
%
% electrodynamics
%

	write "    electrodynamics..." $

	% field strength
	clear f2 $
	pform f2 = 2 $
	f2 := d a1 $
	
	% excitation
	clear h2 $
	pform h2 = 2 $
	h2 := # f2 $
	
	% check of the homogeneous Maxwell equation:
	% this term should vanish
	clear maxhom3 $
	pform mmaxhom3 = 3 $
	maxhom3 := d f2 $
	
	% check of the inhomogeneous Maxwell equation:
	% this term should vanish
	clear maxinhom3 $
	pform mtaxinhom3 = 3 $
	maxinhom3 := d h2 $
	
	% electromagnetic energy-momentum 3-form
	clear sigma3 $
	pform sigma3(a) = 3 $
	sigma3(a) := 1/2 * ( f2 ^ (e(a) _| h2) - h2 ^ (e(a) _| f2) ) $
	
	% trace
	clear trace4 $
	pform trace4 = 4 $
	trace4 := o(-a) ^ sigma3(a) $

	write "        done." $
	
%%%%%%%%%%%%%%%%%%%%%%%%%%%%%%%%%%%%%%%%%%%%%%%%%%
%
% check of Einstein-Maxwell equations
%

	% Einstein-Maxwell equations
	clear emtest3 $
	pform emtest3(a) = 3 $
	emtest3(a) := einstein3(a) + c_lambda*eta3(a) - 2*sigma3(a) $

	end $
\end{lstlisting}
\end{singlespace}

\subsubsection{Decomposition of curvature}
\begin{singlespace}
\begin{lstlisting}[title=file: curvature\_v1.rei]
%%%%%%%%%%%%%%%%%%%%%%%%%%%%%%%%%%%%%%%%%%%%%%%%%
%
% REDUCE file
%
% purpose: calculates components of curvature,
% its decomposition, and duals
%
% last edited by J. Boos, Dec 3, 2014
%
% file: curvature_v1.rei
%
% conventions: 05_elm_inv_v5.pdf
%
%%%%%%%%%%%%%%%%%%%%%%%%%%%%%%%%%%%%%%%%%%%%%%%%%

%%%%%%%%%%%%%%%%%%%%%%%%%%%%%%%%%%%%%%%%%%%%%%%%%%
%
% Riemann curvature
%

	write "    Riemann curvature..." $

	% tensor components
	clear riem0 $
	pform riem0(a, b, c, d) = 0 $
	riem0(c, d, a, b) := e(d) _| (e(c) _| riem2(a, b)) $
	
	% left, right and double dual of Riemann
	clear riemld0, riemrd0, riemdd0 $
	pform {riemld0(a,b,c,d), riemrd0(a,b,c,d), riemdd0(a,b,c,d)} = 0 $
	riemld0(a,b,c,d) := 1/2*eta0(a,b,i,j)*riem0(-i,-j,c,d) $
	riemrd0(a,b,c,d) := 1/2*riem0(a,b,-i,-j)*eta0(i,j,c,d) $
	riemdd0(a,b,c,d) := 1/4*eta0(a,b,i,j)*riem0(-i,-j,-k,-l)*eta0(k,l,c,d) $
	
	write "       done." $

% %%%%%%%%%%%%%%%%%%%%%%%%%%%%%%%%%%%%%%%%%%%%%%%%
%
% Ricci curvature
%

	write "    Ricci curvature..." $	

	% Ricci 1-form
	clear ricci1 $
	pform ricci1(a) = 1 $
	ricci1(a) := e(-b) _| riem2(a, b) $
	
	% Ricci 0-form
	clear ricci0 $
	pform ricci0 = 0 $
	ricci0 := e(-a) _| ricci1(a) $
	
	% traceless Ricci 1-form
	clear tracelessricci1 $
	pform tracelessricci1(a) = 1 $
	tracelessricci1(a) := ricci1(a) - 1/4*o(a)*ricci0 $
	
	write "       done." $
	
%%%%%%%%%%%%%%%%%%%%%%%%%%%%%%%%%%%%%%%%%%%%%%%%%%
%
% irreducible decomposition of curvature
%

	write "    irreducible decomposition of curvature..." $

	% Ricci part of curvature
	clear riccipart2 $
	pform riccipart2(a, b) = 2 $
	riccipart2(a, b) := -1/2*( o(a) ^ tracelessricci1(b) - o(b) ^ tracelessricci1(a)) $
	clear riccipart0 $
	pform riccipart0(a, b, c, d) = 0 $
	riccipart0(c, d, a, b) := e(d) _| ( e(c) _| riccipart2(a, b)) $
	
	% scalar part of curvature
	clear scalarpart2 $
	pform scalarpart2(a, b) = 2 $
	scalarpart2(a, b) := -1/12*ricci0*(o(a) ^ o(b)) $
	clear scalarpart0 $
	pform scalarpart0(a, b, c, d) = 0 $
	scalarpart0(c, d, a, b) := e(d) _| ( e(c) _| scalarpart2(a, b)) $

	% Weyl 2-form
	clear weyl2 $
	pform weyl2(a, b) = 2 $
	weyl2(a, b) := riem2(a, b) - riccipart2(a, b) - scalarpart2(a, b) $
	
	% Weyl anholonomic components
	clear weyl0 $
	pform weyl0(a, b, c, d) = 0 $
	weyl0(c, d, a, b) := e(d) _| (e(c) _| weyl2(a, b)) $
	
	% Weyl dual anholonomic components
	clear weyld0 $
	pform weyld0(a, b, c, d) = 0 $
	weyld0(c, d, a, b) := e(d) _| (e(c) _| ( # weyl2(a, b))) $
	
	% left, right and double dual of Weyl
	clear weylld0, weylrd0, weyldd0 $
	pform {weylld0(a,b,c,d), weylrd0(a,b,c,d), weyldd0(a,b,c,d)} = 0 $
	weylld0(a,b,c,d) := 1/2*eta0(a,b,i,j)*weyl0(-i,-j,c,d) $
	weylrd0(a,b,c,d) := 1/2*weyl0(a,b,-i,-j)*eta0(i,j,c,d) $
	weyldd0(a,b,c,d) := 1/4*eta0(a,b,i,j)*weyl0(-i,-j,-k,-l)*eta0(k,l,c,d) $

	write "       done." $

	end $
\end{lstlisting}
\end{singlespace}

\subsubsection{Kretschmann and Pontryagin invariants}
\begin{singlespace}
\begin{lstlisting}[title=file: invariants\_v1.rei]
%%%%%%%%%%%%%%%%%%%%%%%%%%%%%%%%%%%%%%%%%%%%%%%%%
%
% REDUCE file
%
% purpose: calculates curvature invariants
% (Kretschmann and Pontryagin)
%
% last edited by J. Boos, Dec 3, 2014
%
% file: invariants_v1.rei
%
% conventions: 05_elm_inv_v5.pdf
%
%%%%%%%%%%%%%%%%%%%%%%%%%%%%%%%%%%%%%%%%%%%%%%%%%

%%%%%%%%%%%%%%%%%%%%%%%%%%%%%%%%%%%%%%%%%%%%%%%%%%
%
% quadratic (pseudo-)invariants
%

	write "    Kretschmann invariant..." $

	% Weyl part of Kretschmann 0-form
	clear kretschmannw0 $
	pform kretschmannw0 = 0 $
	kretschmannw0 := - # ( weyl2(-a, -b) ^ (# weyl2(a, b))) $
	
	% traceless Ricci part of Kretschmann 0-form
	clear kretschmannr0 $
	pform kretschmannr0 = 0 $
	kretschmannr0 := - # ( riccipart2(-a, -b) ^ (# riccipart2(a, b))) $
	
	% scalar part of Kretschmann 0-form
	clear kretschmanns0 $
	pform kretschmanns0 = 0 $
	kretschmanns0 := - # ( (# scalarpart2(a, b)) ^ scalarpart2(-a, -b)) $
	
	% Kretschmann 0-form
	clear kretschmann0 $
	pform kretschmann0 = 0 $
	kretschmann0 := - # ( riem2(-a, -b) ^ (# riem2(a, b))) $
	
	write "        done." $
	
	write "    Pontryagin pseudo-invariant..." $
	
	% Weyl part of Pontryagin 0-pseudo-form
	clear pontryaginw0 $
	pform pontryaginw0 = 0 $
	pontryaginw0 := # ( weyl2(-a, -b) ^ weyl2(a, b)) $
	
	% traceless Ricci part of Pontryagin 0-pseudo-form
	clear pontryaginr0 $
	pform pontryaginr0 = 0 $
	pontryaginr0 := # ( riccipart2(-a, -b) ^ riccipart2(a, b)) $
	
	% scalar part of Pontryagin 0-pseudo-form
	clear pontryagins0 $
	pform pontryagins0 = 0 $
	pontryagins0 := # ( scalarpart2(-a, -b) ^ scalarpart2(a, b)) $
	
	% Pontryagin 0-pseudo-form
	clear pontryagin0 $
	pform pontryagin0 = 0 $
	pontryagin0 := # ( weyl2(-a, -b) ^ weyl2(a, b)) $
	
	write "        done." $

	end $
\end{lstlisting}
\end{singlespace}

\subsubsection{Definition of the Weyl tensor components}
\begin{singlespace}
\begin{lstlisting}[title=file: weyl\_def\_v2.rei]
%%%%%%%%%%%%%%%%%%%%%%%%%%%%%%%%%%%%%%%%%%%%%%%%%
%
% REDUCE file
%
% purpose: defines the non-vanishing components
% of the Weyl tensor to facilitate algebraic
% calculations
%
% last edited by J. Boos, Jan 7, 2015
%
% file: weyl_def_v2.rei
%
% conventions: 05_elm_inv_v6.pdf
%
%%%%%%%%%%%%%%%%%%%%%%%%%%%%%%%%%%%%%%%%%%%%%%%%%

%%%%%%%%%%%%%%%%%%%%%%%%%%%%%%%%%%%%%%%%%%%%%%%%%%
%
% symbolic Weyl tensor (saves calculation time)
%

	write "    symbolic Weyl tensor..." $

	% set entries of symbolic Weyl tensor
	% (calculational trick; oly possible if we already know how the Weyl tensor looks like)
	clear symbweyl0 $
	pform symbweyl0(a, b, c, d) = 0 $
	
	for i := 0:3 do for j := 0:3 do for k := 0:3 do for l := 0:3 do
		symbweyl0(-i, -j, -k, -l) := 0 $
	
	symbweyl0(-0, -1, -0, -1) := -2*symb_ee $
	symbweyl0(-0, -1, -1, -0) := -symbweyl0(-0, -1, -0, -1) $
	symbweyl0(-1, -0, -0, -1) := -symbweyl0(-0, -1, -0, -1) $
	symbweyl0(-1, -0, -1, -0) := symbweyl0(-0, -1, -0, -1) $
	
	symbweyl0(-0, -1, -2, -3) := 2*symb_bb $
	symbweyl0(-1, -0, -2, -3) := -symbweyl0(-0, -1, -2, -3) $
	symbweyl0(-0, -1, -3, -2) := -symbweyl0(-0, -1, -2, -3) $
	symbweyl0(-1, -0, -3, -2) := symbweyl0(-0, -1, -2, -3) $
	symbweyl0(-2, -3, -0, -1) := symbweyl0(-0, -1, -2, -3) $
	symbweyl0(-3, -2, -0, -1) := -symbweyl0(-0, -1, -2, -3) $
	symbweyl0(-2, -3, -1, -0) := -symbweyl0(-0, -1, -2, -3) $
	symbweyl0(-3, -2, -1, -0) := symbweyl0(-0, -1, -2, -3) $
	
	symbweyl0(-0, -2, -0, -2) := symb_ee $
	symbweyl0(-2, -0, -0, -2) := -symbweyl0(-0, -2, -0, -2) $
	symbweyl0(-0, -2, -2, -0) := -symbweyl0(-0, -2, -0, -2) $
	symbweyl0(-2, -0, -2, -0) := symbweyl0(-0, -2, -0, -2) $
	
	symbweyl0(-0, -2, -3, -1) := -symb_bb $
	symbweyl0(-2, -0, -3, -1) := -symbweyl0(-0, -2, -3, -1) $
	symbweyl0(-0, -2, -1, -3) := -symbweyl0(-0, -2, -3, -1) $
	symbweyl0(-2, -0, -1, -3) := symbweyl0(-0, -2, -3, -1) $
	symbweyl0(-3, -1, -0, -2) := symbweyl0(-0, -2, -3, -1) $
	symbweyl0(-1, -3, -0, -2) := -symbweyl0(-0, -2, -3, -1) $
	symbweyl0(-3, -1, -2, -0) := -symbweyl0(-0, -2, -3, -1) $
	symbweyl0(-1, -3, -2, -0) := symbweyl0(-0, -2, -3, -1) $
	
	symbweyl0(-0, -3, -0, -3) := symb_ee $
	symbweyl0(-3, -0, -0, -3) := -symbweyl0(-0, -3, -0, -3) $
	symbweyl0(-0, -3, -3, -0) := -symbweyl0(-0, -3, -0, -3) $
	symbweyl0(-3, -0, -3, -0) := symbweyl0(-0, -3, -0, -3) $
	
	symbweyl0(-0, -3, -1, -2) := -symb_bb $
	symbweyl0(-3, -0, -1, -2) := -symbweyl0(-0, -3, -1, -2) $
	symbweyl0(-0, -3, -2, -1) := -symbweyl0(-0, -3, -1, -2) $
	symbweyl0(-3, -0, -2, -1) := symbweyl0(-0, -3, -1, -2) $
	symbweyl0(-1, -2, -0, -3) := symbweyl0(-0, -3, -1, -2) $
	symbweyl0(-2, -1, -0, -3) := -symbweyl0(-0, -3, -1, -2) $
	symbweyl0(-1, -2, -3, -0) := -symbweyl0(-0, -3, -1, -2) $
	symbweyl0(-2, -1, -3, -0) := symbweyl0(-0, -3, -1, -2) $
	
	symbweyl0(-2, -3, -2, -3) := 2*symb_ee $
	symbweyl0(-3, -2, -2, -3) := -symbweyl0(-2, -3, -2, -3) $
	symbweyl0(-2, -3, -3, -2) := -symbweyl0(-2, -3, -2, -3) $
	symbweyl0(-3, -2, -3, -2) := symbweyl0(-2, -3, -2, -3) $
	
	symbweyl0(-3, -1, -3, -1) := -symb_ee $
	symbweyl0(-1, -3, -3, -1) := -symbweyl0(-3, -1, -3, -1) $
	symbweyl0(-3, -1, -1, -3) := -symbweyl0(-3, -1, -3, -1) $
	symbweyl0(-1, -3, -1, -3) := symbweyl0(-3, -1, -3, -1) $
	
	symbweyl0(-1, -2, -1, -2) := -symb_ee $
	symbweyl0(-2, -1, -1, -2) := -symbweyl0(-1, -2, -1, -2) $
	symbweyl0(-1, -2, -2, -1) := -symbweyl0(-1, -2, -1, -2) $
	symbweyl0(-2, -1, -2, -1) := symbweyl0(-1, -2, -1, -2) $
	
	write "        done." $

	end $
\end{lstlisting}
\end{singlespace}

\subsubsection{Newman--Penrose formalism}
\begin{singlespace}
\begin{lstlisting}[title=file: newman\_penrose\_v1.rei]
%%%%%%%%%%%%%%%%%%%%%%%%%%%%%%%%%%%%%%%%%%%%%%%%%
%
% REDUCE file
%
% purpose: defines a complex null tetrad and
% calculates the complex Weyl scalars
%
% last edited by J. Boos, Dec 3, 2014
%
% file: newman_penrose_v1.rei
%
% conventions: 05_elm_inv_v5.pdf
%
%%%%%%%%%%%%%%%%%%%%%%%%%%%%%%%%%%%%%%%%%%%%%%%%%

%%%%%%%%%%%%%%%%%%%%%%%%%%%%%%%%%%%%%%%%%%%%%%%%%%
%
% Newman Penrose formalism
%
	
	write "    setting up complex null tetrad..." $ 
	
	% define complex null tetrad
	% (conventions Senovilla GRG 1997)
	
	% as 1-forms
	clear {np_m1, np_mbar1, np_k1, np_l1} $
	pform np_m1 = 1, np_mbar1 = 1, np_k1 = 1, np_l1 = 1 $
	np_m1    := 1/sqrt(2)*(o(2) - i*o(3)) $
	np_mbar1 := 1/sqrt(2)*(o(2) + i*o(3)) $
	np_k1    := 1/sqrt(2)*(o(0) - o(1)) $
	np_l1    := 1/sqrt(2)*(o(0) + o(1)) $
	
	% components
	clear {np_m0, np_mbar0, np_k0, np_l0} $
	pform np_m0(a) = 0, np_mbar0(a) = 0, np_k0(a) = 0, np_l0(a) = 0 $
	np_m0(a)    := e(a) _| np_m1 $
	np_mbar0(a) := e(a) _| np_mbar1 $
	np_k0(a)    := e(a) _| np_k1 $
	np_l0(a)    := e(a) _| np_l1 $
	
	% Newman-Penrose coefficients for Weyl tensor
	clear {np_psi_0, np_psi_1, np_psi_2, np_psi_3, np_psi_4} $
	pform {np_psi_0, np_psi_1, np_psi_2, np_psi_3, np_psi_4} = 0 $
	
	% transverse wave component in k direction
	np_psi_0 := symbweyl0(-a,-b,-c,-d)*np_l0(a)*np_m0(b)*np_l0(c)*np_m0(d) $
	
	% longitudinal wave component in k direction
	np_psi_1 := symbweyl0(-a,-b,-c,-d)*np_l0(a)*np_k0(b)*np_l0(c)*np_m0(d) $
	
	% "Coulomb" component
	np_psi_2 := 1/2*symbweyl0(-a,-b,-c,-d)*np_l0(a)*np_k0(b)*(np_l0(c)*np_k0(d) - np_m0(c)*np_mbar0(d)) $
	
	% longitudinal wave component in l direction
	np_psi_3 := 1/2*symbweyl0(-a,-b,-c,-d)*np_k0(a)*np_l0(b)*np_k0(c)*np_m0(d) $
	
	% transverse wave component in l direction
	np_psi_4 := 1/2*symbweyl0(-a,-b,-c,-d)*np_k0(a)*np_mbar0(b)*np_k0(c)*np_mbar0(d) $
	
	write "        done." $

	end $
\end{lstlisting}
\end{singlespace}

\subsubsection{Matte decomposition of curvature}
\begin{singlespace}
\begin{lstlisting}[title=file :matte\_v1.rei]
%%%%%%%%%%%%%%%%%%%%%%%%%%%%%%%%%%%%%%%%%%%%%%%%%
%
% REDUCE file
%
% purpose: calculates the E and B parts of the
%          Weyl tensor following Matte (1953)
%
% last edited by J. Boos, Mar 10, 2015
%
% file: matte_v1.rei
%
% conventions: 05_elm_inv_v6.pdf
%
%%%%%%%%%%%%%%%%%%%%%%%%%%%%%%%%%%%%%%%%%%%%%%%%%

%%%%%%%%%%%%%%%%%%%%%%%%%%%%%%%%%%%%%%%%%%%%%%%%%%
%
% E and H tensors
%
	
	write "    Matte's E and B tensors" $
	
	% E Weyl
	clear mattwe0 $
	pform mattwe0(a,b) = 0 $
	mattwe0(a,b) := weyl0(a,0,b,0)/ee $
	
	% B Weyl
	clear mattwb0 $
	pform mattwb0(a,b) = 0 $
	mattwb0(a,b) := -1/2*eta0(a,0,-k,-l)*weyl0(k,l,b,0)/bb $
	
	write "        done." $

	end $
\end{lstlisting}
\end{singlespace}

\subsubsection{Bel tensor}
\begin{singlespace}
\begin{lstlisting}[title=file: bel\_v1.rei]
%%%%%%%%%%%%%%%%%%%%%%%%%%%%%%%%%%%%%%%%%%%%%%%%%
%
% REDUCE file
%
% purpose: calculates the Bel tensor
%
% last edited by J. Boos, Dec 3, 2014
%
% file: bel_v1.rei
%
% conventions: 05_elm_inv_v5.pdf
%
%%%%%%%%%%%%%%%%%%%%%%%%%%%%%%%%%%%%%%%%%%%%%%%%%

%%%%%%%%%%%%%%%%%%%%%%%%%%%%%%%%%%%%%%%%%%%%%%%%%%
%
% Bel tensor
%

	write "    Bel tensor" $

	% definition with duals
	clear robinson0 $
	pform robinson0(a,b,c,d) = 0 $
	robinson0(a,b,c,d) := 1/2*( riem0(a,i,j,c)*riem0(b,-i,-j,d) + riemld0(a,i,j,c)*riemld0(b,-i,-j,d) + riemrd0(a,i,j,c)*riemrd0(b,-i,-j,d) + riemdd0(a,i,j,c)*riemdd0(b,-i,-j,d) ) $
	
	% definition with duals carried out already
	clear robinson20 $
	pform robinson20(a,b,c,d) = 0 $
	robinson20(a,b,c,d) := riem0(a,i,j,c)*riem0(b,-i,-j,d) + riem0(a,i,j,d)*riem0(b,-i,-j,c) - 1/2*( g(a,b)*riem0(i,j,k,c)*riem0(-i,-j,-k,d) + g(c,d)*riem0(i,j,k,a)*riem0(-i,-j,-k,b) ) + 1/8*g(a,b)*g(c,d)*riem0(i,j,k,l)*riem0(-i,-j,-k,-l) $
	
	% this tensor vanishes, since the above two methods are equivalent
	% clear robinsontest0 $
	% pform robinsontest0(a,b,c,d) = 0 $
	% robinsontest0(a,b,c,d) := robinson0(a,b,c,d) - robinson20(a,b,c,d) $
	
	% introduce most general form of energy-momentum like term without fixed summation indices
	clear gen3 $
	pform gen3(a,b,c,d,e) = 3 $
	gen3(k,a,b,c,d) := 1/2*( riem2(a,b) ^ (e(k) _| (# riem2(c,d))) - (# riem2(a,b)) ^ (e(k) _| riem2(c,d)) ) $
	
	% test symmetries of this 3-form
	
	% yields zero: symmetric in b,c
	% robinsontest0(k,a,b,c) := e(k) _| (# gen3(a,b,-j,j,c)) - e(k) _| (# gen3(a,c,-j,j,b)) $
	
	% yields zero: symmetric in a,k
	% robinsontest0(k,a,b,c) := e(k) _| (# gen3(a,b,-j,j,c)) - e(a) _| (# gen3(k,b,-j,j,c)) $
	
	% hypothetical Robinson 3-form
	% does not yield zero: symmetry in ka <-> bc needs to be put in by hand!
	clear robinsonhyp0 $
	pform robinsonhyp0(a,b,c,d) = 0 $
	robinsonhyp0(k,a,b,c) := e(k) _| (# gen3(a,b,-j,j,c)) - e(b) _| (# gen3(c,k,-j,j,a)) $
	
	% build up hypothetical Robinson tensor by means of 3-form above
	clear robinsontilde0 $
	pform robinsontilde0(k,a,b,c) = 0 $
	robinsontilde0(k,a,b,c) := 1/2 * ( e(k) _| (# gen3(a,b,-j,j,c)) + e(b) _| (# gen3(c,k,-j,j,a)) ) $
									 
	% define trace of the above
	clear robinsontrace0 $
	pform robinsontrace0(a,b) = 0 $
	robinsontrace0(a,b) := robinsontilde0(j,-j,a,b) $
	
	% subtract the trace manually
	clear robinson2tilde0 $
	pform robinson2tilde0(k,a,b,c) = 0 $
	robinson2tilde0(k,a,b,c) := 2*( robinsontilde0(k,a,b,c) - 1/4*( g(k,a)*robinsontrace0(b,c) + g(b,c)*robinsontrace0(k,a) ) + 1/16*g(k,a)*g(b,c)*robinsontrace0(i,-i) ) $
	
	% yieldszero: Bel tensor is now indeed traceless in its first and second pair of indices
	clear traceest0 $
	pform tracetest0(a,b) = 0 $
	tracetest0(a,b) := robinson2tilde0(-i,i,a,b) $
	
	% yields zero: Bel tensor can be written as 3-form, with symmetries put in by hand and traces subtracted by hand as well
	clear robinsontest0 $
	pform robinsontest0(a,b,c,d) = 0 $
	robinsontest0(a,b,c,d) := robinson0(a,b,c,d) - robinson2tilde0(a,b,c,d) $
	
	write "        done." $

	end $
\end{lstlisting}
\end{singlespace}

\subsubsection{Bel--Robinson tensor}
\begin{singlespace}
\begin{lstlisting}[title=file: bel\_robinson\_v2.rei]
%%%%%%%%%%%%%%%%%%%%%%%%%%%%%%%%%%%%%%%%%%%%%%%%%
%
% REDUCE file
%
% purpose: calculates the Bel-Robinson tensor
%
% last edited by J. Boos, Jan 7, 2015
%
% file: bel_robinson_v2.rei
%
% conventions: 05_elm_inv_v6.pdf
%
%%%%%%%%%%%%%%%%%%%%%%%%%%%%%%%%%%%%%%%%%%%%%%%%%

%%%%%%%%%%%%%%%%%%%%%%%%%%%%%%%%%%%%%%%%%%%%%%%%%%
%
% Bel-Robinson tensor
%
	
	write "    Bel-Robinson tensor" $
	
	% definition with duals
	clear belrobinson0 $
	pform belrobinson0(a,b,c,d) = 0 $
	belrobinson0(a,b,c,d) := weyl0(a,i,j,c)*weyl0(b,-i,-j,d) + weylld0(a,i,j,c)*weylld0(b,-i,-j,d) $
	
	% definition with duals carried out already
	clear belrobinson20 $
	pform belrobinson20(a,b,c,d) = 0 $
	belrobinson20(a,b,c,d) := weyl0(a,i,j,c)*weyl0(b,-i,-j,d) + weyl0(a,i,j,d)*weyl0(b,-i,-j,c) - 1/2*( g(a,b)*weyl0(i,j,k,c)*weyl0(-i,-j,-k,d) + g(c,d)*weyl0(i,j,k,a)*weyl0(-i,-j,-k,b) ) + 1/8*g(a,b)*g(c,d)*weyl0(i,j,k,l)*weyl0(-i,-j,-k,-l) $
	
	% this tensor vanishes, since the above two methods are equivalent
	% clear belrobinsontest0 $
	% pform belrobinsontest0(a,b,c,d) = 0 $
	% belrobinsontest0(a,b,c,d) := belrobinson0(a,b,c,d) - belrobinson20(a,b,c,d) $
	
	% introduce most general form of energy-momentum like term without fixed summation indices
	clear gen23 $
	pform gen23(a,b,c,d,e) = 3 $
	gen23(k,a,b,c,d) := weyl2(a,b) ^ (e(k) _| (# weyl2(c,d))) - (# weyl2(a,b)) ^ (e(k) _| weyl2(c,d)) $
	
	% the Bel-Robinson 3-form is the following trace of the above
	clear belrobinson3 $
	pform belrobinson3(a,b,c) = 3 $
	belrobinson3(a,b,c) := gen23(a,b,-j,j,c) $
	
	% build up Bel-Robinson tensor by means of 3-form above
	% (this time, no traces need to be subtracted since it is tracefree by design)
	clear belrobinsontilde0 $
	pform belrobinsontilde0(k,a,b,c) = 0 $
	belrobinsontilde0(k,a,b,c) := e(k) _| (# belrobinson3(a,b,c)) $
	
	% check if it is indeed traceless
	clear beltracetest0 $
	pform beltracetest0(a,b) = 0 $
	beltracetest0(a,b) := belrobinson0(-i,a,i,b) $
	
	% this tensor vanishes if the Bel-Robinson tensor can indeed be written in terms of a 3-form
	clear belrobinsontest0 $
	pform belrobinsontest0(a,b,c,d) = 0 $
	belrobinsontest0(a,b,c,d) := belrobinson0(a,b,c,d) - belrobinsontilde0(a,b,c,d) $
	
	% check the Bel-Robinson 3-form against the expression found by Gomez-Lobo
	clear glbelrobinson3 $
	pform glbelrobinson3(a,b,c) = 3 $
	glbelrobinson3(a,b,c) := ( 1/2*g(b,c)*#(weyl2(d,e)^o(a)) + g(e,b)*#(weyl2(c,d)^o(a)) + g(e,c)*#(weyl2(b,d)^o(a)) ) ^ (weyl2(-d,-e)) $

	clear testb3 $
	pform testb3(a,b,c) = 3 $
	testb3(-n,-r,-s) := glbelrobinson3(-n,-r,-s) - (1/2*weyl0(-b,-c,-r,-a)*weyl0(b,c,a,-s)*eta3(-n) - ( weyl0(-n,-c,-r,-a)*weyl0(d,c,a,-s) + weyl0(-n,-c,-s,-a)*weyl0(d,c,a,-r) )*eta3(-d)) $
	
	clear gltest3 $
	pform gltest3(a,b,c) = 3 $
	gltest3(a,b,c) := belrobinson3(a,b,c) - glbelrobinson3(a,b,c) $
	
	% express Bel-Robinson tensor in terms of the complex null tetrad
	clear np_belrobinson0 $
	pform np_belrobinson0(a,b,c,d) = 0 $
	np_belrobinson0(a,b,l,m) := 4*(ee**2+bb**2)*(
													 ( np_m0(a)*np_mbar0(b) + np_m0(b)*np_mbar0(a) + np_l0(a)*np_k0(b) + np_l0(b)*np_k0(a) )
													*( np_m0(l)*np_mbar0(m) + np_m0(m)*np_mbar0(l) + np_l0(l)*np_k0(m) + np_l0(m)*np_k0(l) )
													
													+ ( np_l0(a)*np_m0(b) + np_l0(b)*np_m0(a) ) * ( np_k0(l)*np_mbar0(m) + np_k0(m)*np_mbar0(l) )
													+ ( np_l0(a)*np_mbar0(b) + np_l0(b)*np_mbar0(a) ) * ( np_k0(l)*np_m0(m) + np_k0(m)*np_m0(l) )
													
													+ ( np_k0(a)*np_m0(b) + np_k0(b)*np_m0(a) ) * ( np_l0(l)*np_mbar0(m) + np_l0(m)*np_mbar0(l) )
													+ ( np_k0(a)*np_mbar0(b) + np_k0(b)*np_mbar0(a) ) * ( np_l0(l)*np_m0(m) + np_l0(m)*np_m0(l) )
													
													+ np_l0(a)*np_l0(b)*np_k0(l)*np_k0(m)
													+ np_k0(a)*np_k0(b)*np_l0(l)*np_l0(m)
													+ np_m0(a)*np_m0(b)*np_mbar0(l)*np_mbar0(m)
													+ np_mbar0(a)*np_mbar0(b)*np_m0(l)*np_m0(m)
	
												 ) $
	
	% this tensor vanishes if the above expression is indeed the Bel-Robinson tensor
	clear np_test $
	pform np_test(a,b,c,d) = 0 $
	np_test(a,b,c,d) := belrobinson0(a,b,c,d) - np_belrobinson0(a,b,c,d) $
	
	% define Bonilla's and Senovilla's square root of the Bel-Robinson tensor
	clear sqrt_belrobinson0 $
	pform sqrt_belrobinson0(a,b) = 0 $
	sqrt_belrobinson0(a,b) := 3*sqrt(ee**2+bb**2)*( np_m0(a)*np_mbar0(b) + np_m0(b)*np_mbar0(a) + np_l0(a)*np_k0(b) + np_l0(b)*np_k0(a) ) $
	
	write "        done." $

	end $
\end{lstlisting}
\end{singlespace}

\subsubsection{Kummer tensor}
\begin{singlespace}
\begin{lstlisting}[title=file: kummer\_v1.rei]
%%%%%%%%%%%%%%%%%%%%%%%%%%%%%%%%%%%%%%%%%%%%%%%%%
%
% REDUCE file
%
% purpose: calculates the Kummer tensor
%
% last edited by J. Boos, Dec 3, 2014
%
% file: kummer_v1.rei
%
% conventions: 05_elm_inv_v5.pdf
%
%%%%%%%%%%%%%%%%%%%%%%%%%%%%%%%%%%%%%%%%%%%%%%%%%

%%%%%%%%%%%%%%%%%%%%%%%%%%%%%%%%%%%%%%%%%%%%%%%%%%
%
% Kummer-Weyl tensor (calculate with symbolic Weyl
% tensor to save computation time)
%
	
	write "    Kummer-Weyl tensor" $	
	
	% (symbolic) Kummer Weyl tensor in anholonomic components
	clear symbkummerw0 $
	pform symbkummerw0(a, b, c, d) = 0 $
	symbkummerw0(i, j, k, l) := - symbweyl0(a, i, b, j) * symbweyl0(-a, -c, -b, -d) * symbweyl0(c, k, d, l) $
	
	write "    Kummer-Weyl tensor symbolic irreducible decomposition..." $
	
	write "        1 / 6..." $
	clear symbkummer1w0 $
	pform symbkummer1w0(a, b, c, d) = 0 $
	symbkummer1w0(a, b, c, d) := ( symbkummerw0(a, b, c, d) + symbkummerw0(a, b, d, c) + symbkummerw0(a, d, b, c) + symbkummerw0(a, d, c, b) + symbkummerw0(a, c, d, b) + symbkummerw0(a, c, b, d)
								 + symbkummerw0(c, a, b, d) + symbkummerw0(c, a, d, b) + symbkummerw0(c, d, a, b) + symbkummerw0(c, d, b, a) + symbkummerw0(c, b, d, a) + symbkummerw0(c, b, a, d)
								 + symbkummerw0(b, c, a, d) + symbkummerw0(b, c, d, a) + symbkummerw0(b, d, c, a) + symbkummerw0(b, d, a, c) + symbkummerw0(b, a, d, c) + symbkummerw0(b, a, c, d)
								 + symbkummerw0(d, b, a, c) + symbkummerw0(d, b, c, a) + symbkummerw0(d, c, b, a) + symbkummerw0(d, c, a, b) + symbkummerw0(d, a, c, b) + symbkummerw0(d, a, b, c) ) / 24 $
	
	write "        2 / 6..." $
	clear symbkummer2w0 $
	pform symbkummer2w0(a, b, c, d) = 0 $
	symbkummer2w0(a, b, c, d) := ( symbkummerw0(a, b, c, d) + symbkummerw0(c, b, a, d) - symbkummerw0(b, a, d, c) - symbkummerw0(d, a, b, c) ) / 4 $
	
	write "        3 / 6..." $
	clear symbkummer3w0 $
	pform symbkummer3w0(a, b, c, d) = 0 $
	symbkummer3w0(a, b, c, d) := ( symbkummerw0(a, b, c, d) + symbkummerw0(a, b, d, c) - symbkummerw0(a, d, c, b) - symbkummerw0(a, c, d, b)
							     + symbkummerw0(b, a, c, d) + symbkummerw0(b, a, d, c) - symbkummerw0(b, d, c, a) - symbkummerw0(b, c, d, a) ) / 6 $
	
	write "        4 / 6..." $
	clear symbkummer4w0 $
	pform symbkummer4w0(a, b, c, d) = 0 $
	symbkummer4w0(a, b, c, d) := ( symbkummerw0(a, b, c, d) - symbkummerw0(a, b, d, c) + symbkummerw0(c, b, a, d) - symbkummerw0(c, b, d, a)
		    					 + symbkummerw0(b, a, d, c) - symbkummerw0(b, a, c, d) + symbkummerw0(b, c, d, a) - symbkummerw0(b, c, a, d) ) / 6 $
	
	write "        5 / 6..." $
	clear symbkummer5w0 $
	pform symbkummer5w0(a, b, c, d) = 0 $
	symbkummer5w0(a, b, c, d) := ( symbkummerw0(a, b, c, d) - symbkummerw0(c, b, a, d) - symbkummerw0(b, a, d, c) + symbkummerw0(d, a, b, c) ) / 4 $
	
	write "        6 / 6..." $
	clear symbkummer6w0 $
	pform symbkummer6w0(a, b, c, d) = 0 $
	symbkummer6w0(a, b, c, d) := ( symbkummerw0(a, b, c, d) - symbkummerw0(a, b, d, c) + symbkummerw0(a, d, b, c) - symbkummerw0(a, d, c, b) + symbkummerw0(a, c, d, b) - symbkummerw0(a, c, b, d)
								 + symbkummerw0(c, a, b, d) - symbkummerw0(c, a, d, b) + symbkummerw0(c, d, a, b) - symbkummerw0(c, d, b, a) + symbkummerw0(c, b, d, a) - symbkummerw0(c, b, a, d)
								 + symbkummerw0(b, c, a, d) - symbkummerw0(b, c, d, a) + symbkummerw0(b, d, c, a) - symbkummerw0(b, d, a, c) + symbkummerw0(b, a, d, c) - symbkummerw0(b, a, c, d)
								 + symbkummerw0(d, b, a, c) - symbkummerw0(d, b, c, a) + symbkummerw0(d, c, b, a) - symbkummerw0(d, c, a, b) + symbkummerw0(d, a, c, b) - symbkummerw0(d, a, b, c) ) / 24 $

	write "        done." $

	write "    Kummer-Weyl scalar..." $
	clear symbkws0 $
	pform symbkws0 = 0 $
	symbkws0 := symbkummer1w0(a, -a, b, -b) $
	write "        done." $
		
	write "    Kummer-Weyl axial scalar..." $
	clear symbkwas0 $
	pform symbkwas0 = 0 $
	symbkwas0 := eta0(-a, -b, -c, -d) * symbkummer6w0(a, b, c, d) $
	write "        done." $
	
	% gives zero: decomposition is indeed correct
	% clear symbkummerdectest0 $
	% pform symbkummerdectest0(a,b,c,d) = 0 $
	% symbkummerdectest0(a,b,c,d) := symbkummerw0(a,b,c,d) - symbkummer1w0(a,b,c,d) - symbkummer2w0(a,b,c,d) - symbkummer3w0(a,b,c,d) - symbkummer4w0(a,b,c,d) - symbkummer5w0(a,b,c,d) - symbkummer6w0(a,b,c,d) $
	
	end $
\end{lstlisting}
\end{singlespace}

\end{document}